\documentclass[twocolumn,prb]{revtex4}
\usepackage{graphicx}
\usepackage{bm}
\usepackage{color}
\newcommand{\bk}{{\bf k}}
\newcommand{\bq}{{\bf q}}
\newcommand{\bqt}{\frac{\bf q}2}

\newcommand{\br}{{\bf r}}

\newcommand{\sigmab}{{\overline \sigma}}
\newcommand{\ellbar}{{\overline \ell}}
\begin{document}
\title{Emergence of charge order in a staggered loop-current phase of 
  cuprate high-temperature superconductors }
\author{W. A. Atkinson} \email{billatkinson@trentu.ca}

\affiliation{Department of Physics and Astronomy, Trent University,
  Peterborough Ontario, Canada, K9J7B8} 
\author{A. P. Kampf}
\email{arno.kampf@physik.uni-augsburg.de} 
\author{S. Bulut}
\affiliation{Theoretical
  Physics III, Center for Electronic Correlations and Magnetism,
  Institute of Physics, University of Augsburg, 86135 Augsburg,
  Germany} 
\date{\today}
\begin{abstract}
We study the emergence of charge ordered phases within a $\pi$-loop current ($\pi$LC) model for the 
pseudogap based on a three-band model for underdoped cuprate superconductors. Loop currents and charge 
ordering are driven by distinct components of the short-range Coulomb interactions: loop currents 
result from the repulsion between nearest-neighbor copper and oxygen orbitals, while charge order 
results from repulsion between neighboring oxygen orbitals. We find that the leading $\pi$LC phase 
has an antiferromagnetic pattern similar to previously discovered staggered flux phases, and that it 
emerges abruptly at hole dopings $p$ below the van Hove filling. Subsequent charge ordering 
tendencies in the $\pi$LC phase reveal that diagonal $d$-charge density waves (dCDW) are suppressed 
by the loop currents while axial order competes more weakly. In some cases we find a wide temperature 
range below the loop-current transition, over which the susceptibility towards an axial dCDW is 
large.  In these cases, short-range axial charge order may be induced by doping-related disorder. 
A unique feature of the coexisting dCDW and $\pi$LC phases is the emergence of an incommensurate modulation of the loop currents.  If the dCDW is biaxial (checkerboard) then the resulting incommensurate current pattern breaks all mirror and time-reversal symmetries, thereby
allowing for a polar Kerr effect.
\end{abstract}
\maketitle

\section{Introduction}

Charge order is a universal feature of underdoped cuprate high-temperature superconductors. Charge
ordered phases lie in close proximity to antiferromagnetic, spin-glass, and superconducting phases, 
implying a close competition between the different ordering tendencies.  This raises the possibility 
that some or all of the anomalous properties exhibited by the cuprates are due to multiple competing 
or coexisting electronic phases.  

Originally observed by scanning tunneling spectroscopy in Bi-based 
cuprates\cite{Hoffman:2002bk,Kohsaka:2007hx,Wise:2008}, charge order was then inferred to exist also 
in YBa$_2$Cu$_3$O$_{6+x}$, e.g. from magneto-transport\cite{Daou:2010bo,Chang:2010ic,Chang:2011fw} and 
magneto-oscillation experiments,\cite{Sebastian:2012wh,Harrison:NJP2014} NMR,\cite{Wu:2011ke,Wu:2013} 
and x-ray 
scattering.\cite{Ghiringhelli:2012bw,Chang:2012vf,Blackburn:2013,Blanco-Canosa:2013,Huecker:2014vc} 
More recently, charge order has been found in 
HgBa$_2$CuO$_{4+\delta}$\cite{Doiron:2013,Barisic:2013kz,Tabis:2014kb} and in the electron-doped 
compound Nd$_{2-x}$Ce$_x$CuO$_4$.\cite{daSilvaNeto:2015eq}

The charge order has two distinguishing features:  it has modulation wavevectors $\bq$ that lie 
along the crystalline axes (so-called ``axial order''), and it has an approximate $d_{x^2-y^2}$ 
internal structure.\cite{Kohsaka:2007hx,Mesaros:2011s,Fujita:2014kg,Comin:2014vq,Achkar:2014}  
We therefore adopt the notation $d$-charge density wave (dCDW). In essence, the dCDW can be thought 
of as a predominant charge transfer between neighboring oxygen $p$-orbitals the amplitude of which 
is modulated with wavevector $\bq$.\cite{Fischer:2011,Bulut:2013,Atkinson:2014,Fischer:2014}  

This dCDW is distinct from the stripe order found in La-based cuprates. While both are strongest 
near hole dopings of $p=0.12$, stripes are characterized by an entanglement of spin and charge 
degrees of freedom\cite{Vojta:2009} that is absent in the dCDW phase;\cite{Thampy:2013ty} 
additionally, the doping dependence of the density modulations follows an opposite trend in stripe- 
and charge-ordered materials.\cite{Thampy:2013ty}

Charge order also appears to be distinct from the pseudogap phenomena. Early experiments on  
YBa$_2$Cu$_3$O$_{6+x}$\cite{Ghiringhelli:2012bw,Chang:2012vf}  and 
Bi$_2$Sr$_{2-x}$La$_x$CuO$_{6+x}$\cite{CominScience2014} found that static charge modulations develop 
at temperatures $T_\mathrm{co}$ close to the pseudogap onset temperature $T^\ast$, and this suggested 
a cause for the partial destruction of the Fermi surface that characterizes the pseudogap.  
Furthermore, a recent STM study of Bi$_2$Sr$_2$CaCu$_2$O$_{8+x}$\cite{Hamidian:2015us} found a 
connection between the energy scales of the charge order and the pseudogap. However, systematic 
studies over a wide doping range in YBa$_2$Cu$_3$O$_{6+x}$ have revealed that the onset of the dCDW at 
$T_\mathrm{co}$ varies differently with $p$ than does $T^\ast$.\cite{Blanco-Canosa:2013,Huecker:2014vc}  
In addition, the pseudogap was found insensitive to doping with Zn 
impurities,\cite{Alloul:1991,Zheng:1993,Zheng:1996wh,AlloulRMP:2009} while charge order is rapidly 
quenched.\cite{Blanco-Canosa:2013,Atkinson:2015} Finally, the wavevector associated with the dCDW 
connects tips of the remnant Fermi arcs in the pseudogap phase; this suggests that charge order is 
an instability of, rather than the cause of, the Fermi arcs;\cite{CominScience2014}  indeed, 
theoretical calculations accurately reproduce experimental wavevectors under this 
assumption.\cite{Atkinson:2014,Chowdhury:2014b,Niksic:2015jb,Thomson:2015tg}  

Several calculations found instabilities towards dCDW states with ordering wavevectors $\bq$ oriented 
along the Brillouin zone diagonal (so-called ``diagonal order''),\cite{Metlitski:PRB2010,Metlitski:2010vf,Holder:2012ks,Husemann:2012vg,YamasePRB2012,Efetov:2013,Bulut:2013,Meier:2013,Sau:2013vw}  
in contrast to all the experiments, which find axial order.  This discrepancy is resolved by 
imposing a  pseudogap, from which charge order 
emerges.\cite{Chowdhury:2014,Atkinson:2014,Thomson:2015tg,Feng:2015} This is not a unique 
resolution, though: some authors pointed out that axial and diagonal instabilities are close 
competitors,\cite{Sachdev:2013,Wang:2014wc} and in Ref.~\onlinecite{Yamakawa:2015hb} the inclusion 
of Aslamazov-Larkin vertex corrections led to axial order. Empirically, however, it does appear that 
the pseudogap is a prerequisite for the formation of the dCDW in hole-doped cuprates, since  
$T_\mathrm{co}$ is always less than or equal to $T^\ast$. While the underlying reason is unclear, it 
is possible that short quasiparticle lifetimes at temperatures $T > T^\ast$ inhibit the formation of 
charge order.\cite{Bauer:2015}

If a correct description of the dCDW requires a basic understanding of the pseudogap phase, then 
it is disheartening that the cause of the pseudogap is still unknown.  Many recent proposals suggest 
that the pseudogap is the result of fluctuations of, or competition between, multiple distinct order 
parameters\cite{Wang:2015iq,Wang:2015uw,Hayward:2014,Pepin:2014,Kloss:2015} involving charge and 
superconductivity.   Alternatively, dynamical mean-field calculations find that in the strongly 
correlated limit, local Coulomb interactions may generate a spectral pseudogap without need for a 
true phase transition; this is linked to dynamical antiferromagnetic 
correlations.\cite{Kyung:2006wb,Gunnarsson:2015ia}   However, there is experimental evidence for a 
true thermodynamic phase transition\cite{Shekhter:2013eh,Ramshaw:2015di} at $T^\ast$ (although this
has been challenged in Ref.~\onlinecite{Cooper:2014}) that terminates at a quantum critical point 
near $p=0.19.$\cite{Castellani:1997uk,Varma:1997,Tallon:2001vj,Taillefer:2010gl}  One prominent 
suggestion is that the phase below $T^\ast$ breaks time-reversal symmetry via microscopic loop 
currents (LCs) that may\cite{Marston:1988,Wang:1990,Chakravarty:2001,LeeRMP:2006,Laughlin:2014} or 
may not\cite{Varma:1997} break the translational symmetry of the lattice.

Considerations about the relationship between the dCDW and the pseudogap recently led us to reexamine 
the instabilities of multi-orbital models for cuprate superconductors.\cite{Bulut:2015} For physically 
relevant model parameters, we found a leading instability towards a spontaneous $\pi$-loop current 
($\pi$LC) phase, in which the circulation of the loop currents alternates to form an orbital 
antiferromagnet, similar to staggered LC phases that have been proposed in the 
past.\cite{Marston:1988,Wang:1990,Chakravarty:2001,LeeRMP:2006,Laughlin:2014}  While direct experimental
evidence for staggered LC phases in cuprates is still 
lacking,\cite{Mook:2002,Mook:2002hb,Stock:2002hn,Mook:2004,Sonier:2009fp} we are nonetheless motivated 
to study the LC phase for two reasons: first, the persistence with which LC phases are predicted by 
theory makes it plausible that there exist systems in which LCs are of key importance; second, phase 
competition of the type found in the cuprates can lead to emergent properties that are distinct from 
those of the constituent phases.       

Here, our starting point is the assumption that the pseudogap follows from a $\pi$LC phase, and we 
focus on the possible emergence of charge order within this phase. The required formalism is developed 
in Sec.~\ref{sec:calcs}, and results thereby obtained are presented in Sec.~\ref{sec:results}. We show 
in Sec.~\ref{sec:normal} that the encountered phases originate from  different interactions: the 
$\pi$LC phase is driven primarily by the Coulomb repulsion between nearest-neighbor copper and oxygen 
orbitals, while charge ordering is driven by oxygen-oxygen repulsion. In Sec.~\ref{sec:LC} we discuss 
that axial dCDWs can emerge within the $\pi$LC phase while diagonal dCDWs are strongly suppressed. In 
some cases we find a wide temperature range below the $\pi$LC transition and above the axial dCDW 
transition, 
over which the susceptibility towards an axial dCDW is large.  In these cases, short-range charge 
order may be induced by doping-related disorder. One important consequence relates to the Kerr effect 
that has been measured in YBa$_2$Cu$_3$O$_{6+x}$;\cite{Xia:2008cc,He:2011} a nonzero signal implies that both time-reversal and 
mirror symmetries are broken. The spontaneous currents in the $\pi$LC phase break time-reversal 
symmetry, and mirror symmetries are further broken with the development of the dCDW phase. The 
coexistence of loop currents and dCDW order therefore offers a candidate case for the observed Kerr
rotation.

\section{Calculations}
\label{sec:calcs}
\subsection{Hamiltonian}
\label{sec:Ham}
\begin{figure}
\includegraphics[width=0.5\textwidth]{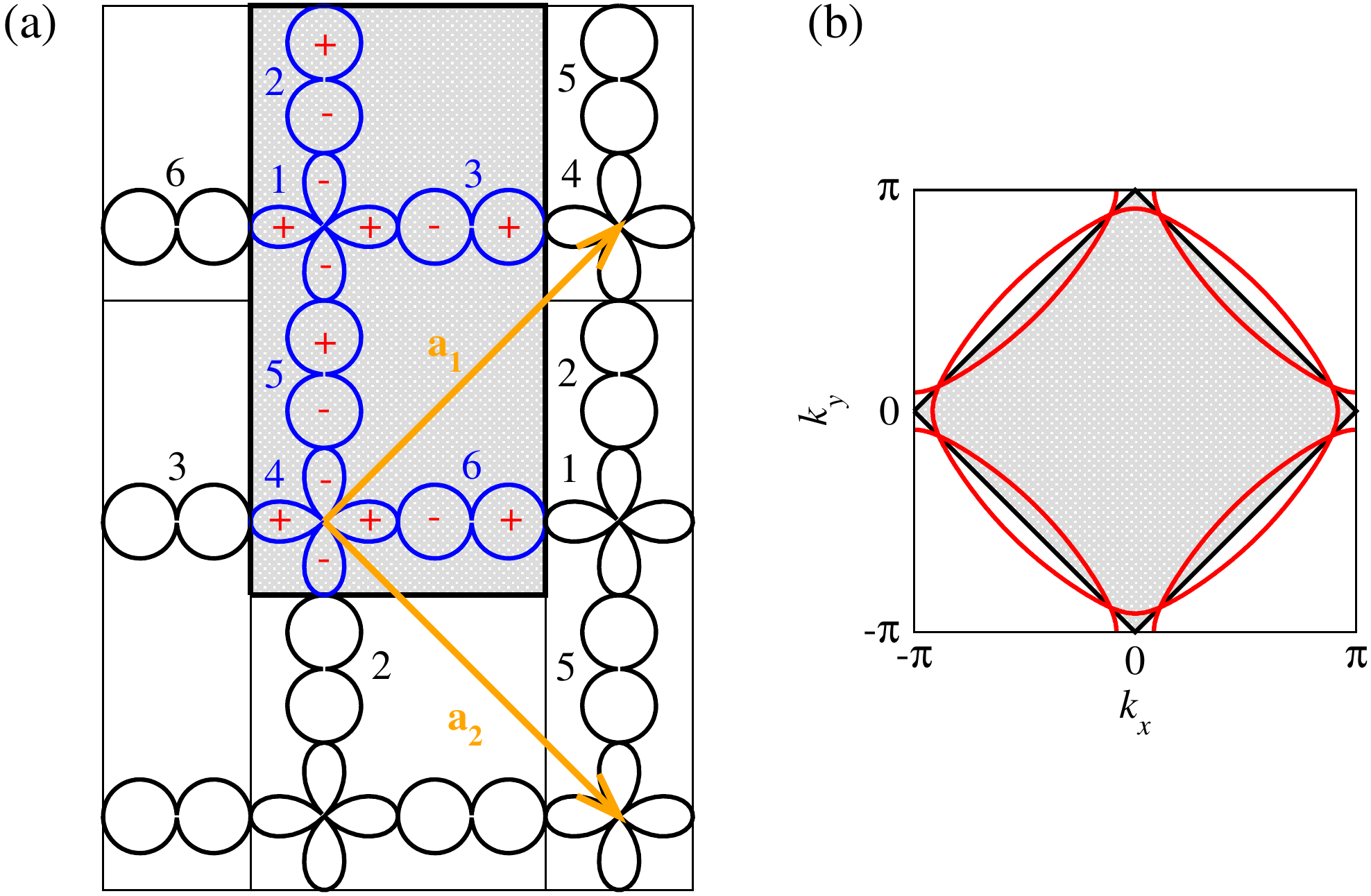}
\caption{(Color online) Unit supercell and Brillouin zone. (a) The 
  supercell (shaded region) contains two CuO$_2$ primitive unit cells, with
  orbitals numbered 1 through 6 as shown.  The plus and minus signs
  indicate the sign convention for the lobes of the Cu$d_{x^2-y^2}$,
  O$p_x$, and O$p_y$ orbitals.  The lattice vectors ${\bf a}_1$ and
  ${\bf a}_2$ lead to the folded Brillouin zone shown in (b) (shaded
  region). Backfolded Fermi surfaces are shown for hole filling
  $p=0.15$ (thick red lines).  }
\label{fig:unitcell}
\end{figure}

We adopt a three-band model for the CuO$_2$ primitive unit cell, as described in 
Ref.~\onlinecite{Bulut:2013}.  The model includes the
Cu$d_{x^2-y^2}$ orbital and the O$p$ orbital from each oxygen that
forms a $\sigma$ bond with it; we label these O$p_x$ and O$p_y$.  The
noninteracting part of the Hamiltonian is 
\begin{equation}
\hat H_0 = \sum_{\bk,\sigma} \sum_{\alpha,\beta} c^\dagger_{\bk\alpha\sigma}
h_{0,\alpha\beta}(\bk) c_{\bk\beta\sigma},
\label{eq:bareham}
\end{equation}
where $\sigma$ is a spin index and $\alpha$, $\beta$ denote the orbitals.
We take the convention that $c_{\bk \beta\sigma}$ is an {\em
  electron} annihilation operator.  Because the $\pi$LC phase has a
periodicity of two unit cells, we use a supercell comprising two
primitive CuO$_2$ unit cells so that orbital labels run from 1 to 6
(Fig.~\ref{fig:unitcell}).

We assume that the SU(2) spin invariance is unbroken so that spin-up
and spin-down electrons satisfy identical equations of motion.  For
brevity, we therefore suppress the spin-index except where it is
required.

The Hamiltonian has diagonal matrix elements $h_{0,\alpha\alpha}(\bk)$
given by the on-site energies $\epsilon_d$ (for $\alpha = 1,4$) and
$\epsilon_p$ (otherwise).  The model further includes nearest-neighbor
hopping between Cu and O orbitals with amplitude $t_{pd}$, and
between adjacent O orbitals with amplitude $t_{pp}$.  The Hamiltonian
matrix in Eq.~(\ref{eq:bareham}) is therefore
\begin{equation}
{\bf h}_0(\bk) = \left [ \begin{array}{cc} 
{\bf h}_1(\bk) & {\bf h}_2(\bk)\\
{\bf h}_2(\bk)^\dagger & {\bf h}_1(\bk)
\end{array} \right ],
\end{equation}
where
\begin{eqnarray}
{\bf h}_1(\bk) &=& \left [ \begin{array}{ccc}
 \epsilon_d & t_{pd} e^{ik_y/2} & -t_{pd} e^{ik_x/2}  \\
 t_{pd}e^{-ik_y/2} &  \epsilon_p & 2 t_{pp}c_-  \\
-t_{pd}e^{-ik_x/2} & 2t_{pp}c_- &  \epsilon_p 
\end{array} \right  ], \\
{\bf h}_2(\bk) &=& \left [ \begin{array}{ccc}
 0 & -t_{pd} e^{-ik_y/2} & t_{pd} e^{-ik_x/2} \\ 
 -t_{pd} e^{ik_y/2} & 0 & -2 t_{pp}c_+  \\
 t_{pd} e^{ik_x/2} & -2 t_{pp} c_+ & 0
\end{array} \right ].
\end{eqnarray}
The primitive lattice constant is $a_0 = 1$, and 
$c_\pm = \cos(\frac{k_x}{2} \pm \frac{k_y}{2})$.  The signs of
the off-diagonal matrix elements $h_{0,\alpha\beta}(\bk)$ are
determined by the product of signs of the closest lobes of orbitals
$\alpha$ and $\beta$, as shown in Fig.~\ref{fig:unitcell}(a).  Because
the supercell contains two primitive unit cells, the Brillouin zone is
halved and the Fermi surface is folded into the reduced Brillouin zone
[Fig.~\ref{fig:unitcell}(b)].

We consider both on-site and nearest-neighbor Coulomb repulsion, so
the interaction has the form
\begin{equation}
\hat V = \sum_{i\alpha} U_{\alpha} \hat n_{i\alpha \uparrow} \hat
n_{i\alpha\downarrow} + \frac 12 \sum_{i\alpha \sigma,j\beta\sigma'}
V_{i\alpha,j\beta} \hat n_{i\alpha\sigma} \hat n_{j\beta\sigma'},
\end{equation}
where $i$ and $j$ label supercells, $\alpha$ and $\beta$ label orbitals, 
$\sigma$ and $\sigma'$ label spins, and $\hat n_{i\alpha\sigma} = c^\dagger_{i\alpha\sigma}
c_{i\alpha\sigma}$.  The on-site Coulomb
interaction $U_\alpha$ is $U_d$ ($\alpha = 1,4$) or $U_p$ (otherwise);
the nonlocal interaction $V_{i\alpha,j\beta}$ is $V_{pd}$ for
nearest-neighbor $p$ and $d$ orbitals, and $V_{pp}$ for adjacent
oxygen orbitals.

Following Ref.~\onlinecite{Bulut:2013} we take $t_{pd}=1$ to be the
unit of energy, $t_{pp} = -0.5$, and $\epsilon_d-\epsilon_p=2.5$.  The
interaction strengths are $U_d = 9.0$, $U_p = 3.0$, $V_{pp}$ varies between 
1.0 and 1.3 and $V_{pd}$ between 2.0 and 3.0.

\subsection{Hartree-Fock Approximation}
Interactions are first treated within a Hartree-Fock (HF)
approximation,  $\hat V \approx \hat V_{HF} \equiv \hat V_H + \hat
V_X$, where the Hartree term is
\begin{equation}
\hat V_H = \sum_{i\alpha\sigma} U_{\alpha} \hat n_{i\alpha \sigma} n_{i\alpha\sigmab}
+ \sum_{i\alpha \sigma,j\beta\sigma'} V_{i\alpha,j\beta} \hat n_{i\alpha\sigma} n_{j\beta\sigma'},
\end{equation}
with $\sigmab \equiv -\sigma$ and $n_{i\alpha\sigma} \equiv \langle \hat n_{i\alpha\sigma}\rangle$, 
and the exchange term is
\begin{eqnarray}
  \hat V_X 
 %&=& \frac 12 \sum_{i\alpha\sigma,j\beta\sigma'}
 % V_{i\alpha,j\beta} \hat n_{i\alpha\sigma} \hat n_{j\beta\sigma'} \\ 
%
  &=& - \sum_{i\alpha,j\beta} \sum_\sigma
  c^\dagger_{i\alpha\sigma} c_{j\beta\sigma} V_{i\alpha,j\beta}
  \langle c^\dagger_{j\beta\sigma} c_{i\alpha\sigma} \rangle.
\label{eq:VX}
\end{eqnarray}
Within the HF approximation, the leading instability is to a spin-density wave (SDW)
state involving spins on the Cu sites.\cite{Bulut:2013}  
This state is driven by the large local Coulomb interaction
$U_d$; it is well known that strong correlations suppress the SDW except near half-filling, and
we therefore make a restricted HF approximation that preserves the SU(2) invariance of the spins.
SU(2) symmetry implies $n_{i\alpha \uparrow} = n_{i\alpha \downarrow}$ and 
$\langle c^\dagger_{j\beta\uparrow} c_{i\alpha\uparrow} \rangle
= \langle c^\dagger_{j\beta\downarrow} c_{i\alpha\downarrow} \rangle$,
so that the HF Hamiltonian is identical for spin-up and spin-down electrons.

Expressing $\hat V_{HF}$ in terms of Bloch states (and suppressing the spin index) gives
%\begin{eqnarray}
%\hat V_{HF}&=& \frac 1N \sum_{\bk,\bq,\bk'} \sum_{\alpha,\beta}\sum_{\mu,\nu}
%c^\dagger_{\bk+\bqt\alpha}c_{\bk-\bqt\beta} \nonumber \\
%&&\times \big \{ \left [ U_\alpha\delta_{\alpha,\mu}
%+ 2V_{\alpha\mu} (\bq)\right ] \delta_{\alpha,\beta}\delta_{\mu,\nu} \nonumber \\
%&&- V_{\alpha\beta}(\bk-\bk') \delta_{\nu,\beta}\delta_{\mu,\alpha} \big \}
%\langle c^\dagger_{\bk'-\bqt\nu}c_{\bk'+\bqt\mu} \rangle, 
%\label{eq:VHF}
%\end{eqnarray}
\begin{eqnarray}
\hat V_{HF} &=&\sum_{\bk,\bq, \alpha,\beta} P_{\alpha\beta}(\bk,\bq)
c^\dagger_{\bk+\bqt\alpha} c_{\bk-\bqt\beta},
\label{eq:VHF2}
\end{eqnarray}
where 
\begin{eqnarray}
P_{\alpha\beta}(\bk,\bq) &=& 
\frac{1}{N}\sum_{\bk'} \sum_{\mu,\nu}  
\big \{ \left [ U_\alpha\delta_{\alpha,\mu}
+ 2V_{\alpha\mu} (\bq)\right ] \delta_{\alpha,\beta}\delta_{\mu,\nu}
\nonumber \\
&&- V_{\alpha\beta}(\bk-\bk') \delta_{\nu,\beta}\delta_{\mu,\alpha} \big \}
\langle c^\dagger_{\bk'-\bqt\nu}c_{\bk'+\bqt\mu} \rangle,  \nonumber \\
\label{eq:P}
\end{eqnarray}
is the HF ``self-energy'' and  
\begin{equation}
V_{\alpha\beta}(\bq) = \sum_{\br_{\alpha\beta}} e^{i\bq\cdot
  \br_{\alpha\beta}} V_{\alpha\beta}(\br_{\alpha\beta}),
\label{eq:Vabq}
\end{equation}
with $\{ \br_{\alpha\beta} \}$ the set of intra- and inter-supercell
vectors pointing {\em from} orbital $\alpha$ {\em to} nearest-neighbor
orbital $\beta$.  Explicit expressions for $V_{\alpha\beta}(\bq)$ are
given in Appendix~\ref{app:A}.

In the HF approximation, terms proportional to $U_{\alpha}$
contribute only Hartree terms, while the nonlocal terms make both
Hartree and exchange contributions.  Because our model parameters are
chosen phenomenologically to reproduce the cuprate band structure, the
homogeneous components of the Hartree and exchange self-energies are
 implicitly  present in the site energies $\epsilon_d$
and $\epsilon_p$ and hopping matrix elements $t_{pd}$ and $t_{pp}$.
To avoid double-counting, we retain only the spatially inhomogeneous
components of the interaction self-energy; these will prove responsible for
both loop currents and charge order.

It is convenient to decompose the interactions in Eq.~(\ref{eq:P}) in
a set of basis functions $g^\ell_{\alpha\beta}(\bk)$:
\begin{eqnarray}
\left [ U_\alpha\delta_{\alpha,\mu}
+ 2V_{\alpha\mu} (\bq)\right ] \delta_{\alpha,\beta}\delta_{\mu,\nu} 
- V_{\alpha\beta}(\bk-\bk') \delta_{\nu,\beta}\delta_{\mu,\alpha} 
\nonumber \\
=
\sum_{\ell,\ell'} \tilde V^{\ell\ell'}(\bq) g^\ell_{\alpha\beta}(\bk) g^{\ell'}_{\mu\nu}(\bk')^\ast. \nonumber \\
\label{eq:Vdecomp}
\end{eqnarray}
$g^\ell_{\alpha\beta}(\bk)$ are $6\times 6$ matrices in the orbital indices $\alpha$ and $\beta$, with 
a single nonzero matrix element corresponding to a unique bond or site: 
\begin{equation}
g^\ell_{\alpha\beta}(\bk) = e^{i\bk\cdot \br_{\alpha\beta}} \delta_{\alpha,\alpha_\ell} \delta_{\beta,\beta_\ell},
\end{equation}
where each $\ell$ labels  either a  directed bond pointing from $\alpha_\ell$ to $\beta_\ell$,
or an orbital when $\alpha_{\ell}=\beta_\ell$.  There are a total of 38 orbital pairs 
$(\alpha_\ell,\beta_\ell)$, and these are listed in Table~\ref{tab:P} in
Appendix~\ref{app:A}, along with the corresponding basis functions. Here, we note that 
$\ell \in [1,32]$ labels the directed bonds between nearest-neighbor sites, and 
$\ell \in [33,38]$ labels the six orbitals making up the supercell.  

With the decomposition (\ref{eq:Vdecomp}), we obtain
\begin{eqnarray}
P_{\alpha\beta}(\bk,\bq) 
%&=& \sum_{\ell,\ell'} \tilde
%V^{\ell\ell'}(\bq) g^\ell_{\alpha\beta}(\bk) \nonumber \\
%&&\times \frac{1}{N}\sum_{\bk'}
%\sum_{\mu,\nu} g^{\ell'}_{\mu\nu}(\bk')^\ast \langle
%c^\dagger_{\bk'-\bqt\nu}c_{\bk'+\bqt\mu} \rangle \nonumber \\ 
&=& \sum_{\ell} \tilde P^\ell(\bq) g^\ell_{\alpha\beta}(\bk),
\label{eq:P2}
\end{eqnarray}
where
\begin{equation}
\tilde P^\ell(\bq) = \frac{1}{N}\sum_{\bk'} \sum_{\ell',\mu,\nu}
\tilde V^{\ell\ell'}(\bq) g^{\ell'}_{\mu\nu}(\bk')^\ast \langle
c^\dagger_{\bk'-\bqt\nu}c_{\bk'+\bqt\mu} \rangle,
\label{eq:Ptilde}
\end{equation}
is the self-consistency equation for the HF self-energy for bond $\ell$. To perform an unbiased 
search for broken-symmetry phases within HF theory, it is most convenient to linearize 
Eq.~(\ref{eq:Ptilde}) so that it acquires the form
\begin{equation}
\tilde P^\ell(\bq) = -\sum_{\ell',\ell''} \tilde V^{\ell\ell'}(\bq) \tilde X_0^{\ell'\ell''}(\bq) 
\tilde P^{\ell''}(\bq).
\label{eq:linearized}
\end{equation}
This step is performed explicitly in the next section.

\subsection{Linearized Hartree-Fock Equations}

We define a generalized susceptibility that describes the change in
$\tilde P^\ell(\bq)$ induced by a perturbing field
$\tilde \phi^{\ell'}(\bq,t)$, where $\ell$ and $\ell'$ label either bonds or sites as 
described above. In the limit of a vanishingly weak perturbation, a phase transition is 
signalled by a diverging susceptibility eigenvalue.

The general form of the perturbation is
\begin{eqnarray}
\hat \Phi(t) &=& \sum_{m\mu,n\nu} \phi_{m\mu,n\nu}(t) 
c^\dagger_{m\mu} c_{n\nu} \nonumber \\
&=& \sum_{\bk,\bq} \sum_{\mu\nu}  \phi_{\mu\nu}(\bk,\bq,t) 
c^\dagger_{\bk+\bqt \mu} c_{\bk-\bqt\nu}, 
\label{eq:Phit}
\end{eqnarray}
where $m,n$ label supercells and 
\begin{equation}
%\phi_{m\mu,n\nu}(t) &=&
%\frac{1}{N} \sum_{\bk,\bq} \phi_{\mu\nu}(\bk,\bq,t)
%e^{i\bk\cdot(\br_{m\mu}-\br_{n\nu})} e^{i\bqt\cdot (\br_{m\mu}+\br_{n\nu})}.\nonumber \\
\phi_{\mu\nu}(\bk,\bq,t) = \frac{1}{N} \sum_{m,n} 
\phi_{m\mu,n\nu}(t)
e^{i\bk\cdot(\br_{n\nu}-\br_{m\mu})} e^{-i\bqt\cdot (\br_{m\mu}+\br_{n\nu})}. 
\label{eq:PhiFT}
\end{equation}
In this equation, $\bk$ is associated with the relative coordinate
connecting orbitals $\mu$ and $\nu$, while $\bq$ is associated with
the spatial modulation of the field;  a conventional electrostatic
potential would have
\begin{equation}
\phi_{\mu\nu}(\bk,\bq,t) = \delta_{\mu,\nu} \phi_\mu(\bq,t). 
\end{equation}
%Inverting Eq.~(\ref{eq:PhiFT}), we obtain
Provided the perturbation is restricted to on-site and nearest-neighbor
terms, Eq.~(\ref{eq:PhiFT}) can be decomposed in terms of $g^\ell_{\mu\nu}(\bk)$,
\begin{eqnarray}
\phi_{\mu\nu}(\bk,\bq,t) 
%&=& 
%\frac 1N \sum_{\br_{\mu\nu}} e^{i\bk\cdot\br_{\mu\nu}} \phi_{\mu\nu}(\br_{\mu\nu},\bq,t) \nonumber \\
&=& \sum_\ell \tilde \phi^\ell(\bq,t) g^\ell_{\mu\nu}(\bk).
\end{eqnarray}
 Then
\begin{equation}
\hat \Phi(t) = \sum_{\bk,\bq} \sum_{\mu\nu} \sum_\ell 
\tilde \phi^\ell(\bq,t) g^\ell_{\mu\nu}(\bk)
c^\dagger_{\bk+\bqt \mu} c_{\bk-\bqt\nu}.
\end{equation}
Hermiticity of $\hat \Phi(t)$ requires for the perturbing fields
\begin{equation}
\tilde \phi^\ell(-\bq,t) = \tilde \phi^{\ellbar}(\bq,t)^\ast,
\label{eq:hermiticity}
\end{equation}
where $\ell$ and $\ellbar$ describe the same bond, but oriented in
opposite directions.

The perturbing field induces time-dependent collective excitations
$\delta P_{\alpha\beta}(\bk,\bq,t)$ of the self-energy; these feed back into
the linear response, so that the total perturbation is
\begin{eqnarray}
\hat H'(t) &=& \sum_{\bq}\sum_{\ell} \left [ \delta \tilde P^{\ell}(\bq,t) +
  \tilde \phi^{\ell}(\bq,t) \right ] \nonumber \\ &&\times
\sum_{\bk\mu\nu}g^{\ell}_{\mu\nu}(\bk)
c^\dagger_{\bk+\bqt\mu} c_{\bk-\bqt\nu},
\end{eqnarray}
where we have expanded  $\delta P_{\mu\nu}(\bk,\bq,t) = \sum_\ell g^\ell_{\mu\nu}(\bk)
\delta \tilde P^{\ell}(\bq,t)$.

A self-consistent expression for $\delta \tilde P^{\ell}(\bq,t)$ is obtained from 
Kubo's equation for the first order response of the charge density to
$\hat H'(t)$:
\begin{equation}
\delta \tilde P^\ell(\bq,t) = -i \int_{-\infty}^{t} dt' 
\langle [ \hat P^\ell(\bq,t), \hat H'(t')]\rangle,
\end{equation}
where
\begin{equation}
\hat P^\ell(\bq) = \frac{1}{N}\sum_{\bk'} \sum_{\ell',\mu,\nu}
\tilde V^{\ell\ell'}(\bq) g^{\ell'}_{\mu\nu}(\bk')^\ast 
c^\dagger_{\bk'-\bqt\nu}c_{\bk'+\bqt\mu} 
\label{eq:Phat}
\end{equation}
is the operator form of $\tilde P^\ell(\bq)$ [see Eq.~(\ref{eq:Ptilde})].  A straightforward 
calculation yields
\begin{equation}
\delta \mathbf{ \tilde P} (\bq,\omega) = - \mathbf{ \tilde V }(\bq) \mathbf{ \tilde X_0} (\bq,\omega) 
\left [ \delta \mathbf{ \tilde P} (\bq,\omega) + \bm{ \tilde \phi} (\bq,\omega) \right ], 
\label{eq:kubo}
\end{equation}
where bold symbols represent matrices and vectors in the $38\times 38$ bond and orbital basis.
The bare susceptibility matrix has elements
\begin{eqnarray}
\tilde X_0^{\ell\ell'}(\bq,\omega) 
&=& \frac {1}{N} \sum_{\bk} \sum_{\alpha\beta\mu\nu} 
g^\ell_{\mu\nu}(\bk)^\ast g^{\ell'}_{\alpha\beta}(\bk) \nonumber \\
&&\times \sum_{n,n'}
\Psi_{\mu n}(\bk_+) \Psi^\ast_{\alpha n}(\bk_+) 
\Psi_{\beta n'}(\bk_-) \Psi_{\nu n'}^\ast(\bk_-) \nonumber \\
&&\times 
\frac{f(E_{n\bk_+}) - f(E_{n'\bk_-})}
{\omega + i\delta - E_{n\bk_+}+E_{n'\bk_-}},
\label{eq:X0}
\end{eqnarray}
where $\bk_\pm \equiv \bk\pm \bqt$, greek symbols are orbital labels, $n$ and $n'$ are band indices, 
and $E_{n\bk}$ and $\Psi_{\mu n}(\bk)$ are  respectively the eigenvalues and eigenvectors of the 
Hamiltonian ${\hat H}_0$. In the static limit $\omega\rightarrow 0$ and for a vanishingly weak 
external potential $\tilde \phi^{\ell}(\bq,\omega)$, Eq.~(\ref{eq:kubo}) reduces to
Eq.~(\ref{eq:linearized}).

Equation~(\ref{eq:kubo}) is a $38\times 38$ matrix equation that can
be inverted for each $\bq$ and $\omega$ to obtain
\begin{equation}
\delta \mathbf{\tilde P}(\bq,\omega) = -\mathbf{\tilde V}(\bq) 
\mathbf{\tilde X}(\bq,\omega) \bm{\tilde \phi}(\bq,\omega)
\label{eq:PXphi}
\end{equation}
with
\begin{equation}
\mathbf{\tilde X}(\bq,\omega) = \left [\mathbf{1} + 
\mathbf{\tilde X_0} (\bq,\omega) \mathbf{\tilde V}(\bq) \right ]^{-1} 
\mathbf{\tilde X_0}(\bq,\omega).
\end{equation}
Equation~(\ref{eq:PXphi}) describes the change in the HF
self-energy induced by a weak perturbing field.

\subsection{Connection to Charge and Current Densities}
We denote by $\chi(\bq)$ the largest eigenvalue of the static
susceptibility matrix ${\bf \tilde X}(\bq,0)$.
The divergence of  $\chi(\bq)$ as temperature is lowered
signals a phase transition. Further information about the resulting phase is obtained from the
corresponding eigenvector $\mathbf{\tilde v}_\bq$.  In particular, 
both the current and charge density can be obtained from a generalized
charge density,
\begin{equation}
\rho_{i\alpha,j\beta} = \langle
c^\dagger_{i\alpha} c_{j\beta} \rangle,
\end{equation}
which is closely related to the HF self-energy by
\begin{equation}
P_{i\alpha,j\beta} = V_{i\alpha,j\beta} \rho_{i\alpha,j\beta}^\ast.
\label{eq:PVrho}
\end{equation}
For $(i\alpha) = (j\beta)$,
$\rho_{i\alpha,j\beta}$ reduces to the single-spin charge
density $n_{i\alpha}$, while for nearest-neighbor pairs $(i\alpha)$ and
$(j\beta)$, the imaginary part of $\rho_{i\alpha,j\beta}$ gives the
 probability current along the bond from $(i\alpha)$ to $(j\beta)$,
\begin{equation}
J_{i\alpha,j\beta} = -2t_{i\alpha,j\beta} \mbox{Im}[\rho_{i\alpha,j\beta}].
\label{eq:Jrho}
\end{equation}
In Eq.~(\ref{eq:Jrho}), $t_{i\alpha,j\beta}$ is $\pm t_{pd}$ or $\pm
t_{pp}$, depending on the bond type, where the sign depends on the
relative signs of the closest lobes of orbitals $\alpha$ and $\beta$
in Fig.~\ref{fig:unitcell} (thus $t_{i1,i3} = -t_{pd}$; $t_{i5,i6} =
+t_{pp}$).

By Fourier transforming Eq.~(\ref{eq:PVrho}) and expanding left and right sides in terms of 
the basis functions $g^\ell_{\alpha\beta}(\bk)$, we obtain
\begin{equation}
\tilde P^\ell(\bq) = \sum_{\ell'} \tilde V^{\ell\ell'} \tilde \rho^{\ell'}(-\bq)^\ast,
\label{eq:Prho}
\end{equation}
with 
\begin{equation}
\tilde \rho^{\ell'}(\bq) = \frac{1}{N}\sum_{\bk,\alpha,\beta} g^\ell_{\alpha\beta}(\bk) 
\langle c^\dagger_{\bk-\bqt \alpha} c_{\bk+\bqt \beta}\rangle.
\end{equation}
Equation (\ref{eq:Prho}) provides a connection between the induced self energy 
$\delta {\bf \tilde P}(\bq,\omega)$ in Eq.~(\ref{eq:PXphi}) and the corresponding induced 
change in the generalized charge density $\delta \bm{\tilde \rho}(\bq,\omega)$.

Near the phase transition,  the static susceptibility matrix ${\bf \tilde X}(\bq)$ is dominated 
by the diverging  eigenvalue $\chi(\bq)$, such that
\begin{equation}
{\bf \tilde X}(\bq) \approx \chi(\bq) \mathbf{\tilde v}_\bq
\mathbf{\tilde v}_\bq^\dagger,
\end{equation}
where $\mathbf{\tilde v}_\bq$ is the column eigenvector  corresponding to $\chi(\bq)$,
$\mathbf{\tilde v}_\bq^\dagger$ is the transpose conjugate, and the outer product 
$\mathbf{\tilde v}_\bq \mathbf{\tilde v}_\bq^\dagger$ generates a matrix.  Substitution of 
Eq.~(\ref{eq:Prho}) into Eq.~(\ref{eq:PXphi}) immediately yields the induced static 
(generalized) charge density,
\begin{eqnarray}
\delta \bm{\tilde \rho}(-\bq)^\ast &=& -\mathbf{\tilde X}(\bq) \bm{\tilde \phi}(\bq) \nonumber \\
&=& - \varphi_\bq \chi(\bq) \mathbf{\tilde v}_{\bq},
\label{eq:rhoq1}
\end{eqnarray}
where $\varphi_\bq =  \mathbf{\tilde v}_{\bq}^\dagger \cdot \bm{\tilde \phi}(\bq) $ is the
projection of the field onto the diverging eigenmode.  The hermiticity 
condition~(\ref{eq:hermiticity}), along with a similar condition for
${\bf \tilde v}_\bq$ (see Eq.~(\ref{eq:vtransform}) in Appendix~\ref{app:C}), 
imposes the constraint $\varphi_{-\bq} = \varphi_\bq$.   Then,
\begin{eqnarray}
\delta \rho_{i\alpha,j\beta}  &=& \delta \rho(\bq) e^{i\bqt\cdot (\br_{i\alpha}+\br_{j\beta})}
+ \delta \rho(-\bq) e^{-i\bqt\cdot (\br_{i\alpha}+\br_{j\beta}) }\nonumber \\
&=& -\chi(\bq) \left \{ 
 e^{i\bqt\cdot (\br_{i\alpha}+\br_{j\beta})}  
 \varphi_\bq \tilde v_\bq^\ellbar  \right . \nonumber \\
&& \left . + e^{-i\bqt\cdot (\br_{i\alpha}+\br_{j\beta})} 
  \varphi_\bq^{\ast}  \tilde v_\bq^{\ell\ast } \right \},
  \label{eq:drho2}
\end{eqnarray}
where $\ell$ denotes the directed bond from $(i\alpha)$ to $(j\beta)$, and 
and $\ellbar$ denotes the oppositely directed bond.  The complex phase of $\phi_\bq$ 
shifts the density wave spatially, and can therefore be set to zero without loss of generality: 
\begin{eqnarray}
\delta \rho_{i\alpha,j\beta}  &=& -\chi(\bq) \varphi_\bq \left \{ 
 e^{i\bqt\cdot (\br_{i\alpha}+\br_{j\beta})}  
  \tilde v_\bq^\ellbar  \right . \nonumber \\
&& \left . + e^{-i\bqt\cdot (\br_{i\alpha}+\br_{j\beta})} 
  \tilde v_\bq^{\ell\ast } \right \},
  \label{eq:drho}
\end{eqnarray}
Real-space patterns shown in the next section are calculated from the portion of 
Eq.~(\ref{eq:drho}) contained in braces.

%is the generalized charge susceptibility at wavevector $\bq$.
%In the disordered phase the susceptibility is identical to that studied in
%Refs.~\onlinecite{Bulut:2013} and \onlinecite{Bulut:2015}.  There, we showed that the
%susceptibility can be understood as a sum of mixed ladder and bubble
%diagrams.  

\begin{figure}
\includegraphics[width=\columnwidth]{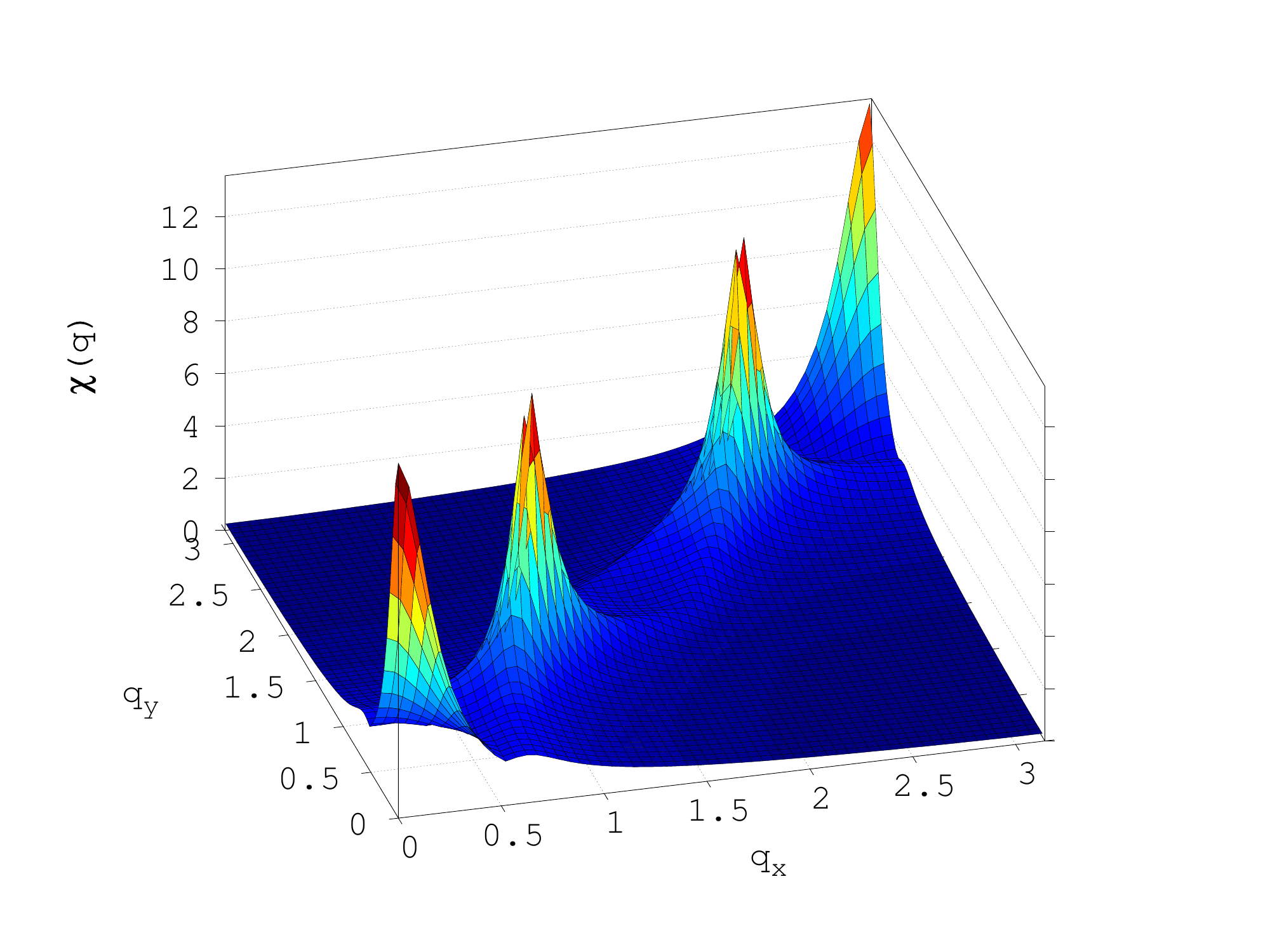}
\caption{(Color online) Largest eigenvalue of the static susceptibility matrix
  $\mathbf{\tilde X}(\bq)$ in the normal state at
  $T=0.010$, $V_{pd}=2.5$, and $p=0.10$.   }
\label{fig:chi0}
\end{figure}

\section{Results}
\label{sec:results}
\subsection{Instabilities of the Normal State}
\label{sec:normal}
As there are no broken symmetries in the normal high-temperature phase, the HF self-energy generates 
only a homogeneous renormalization of the model parameters.  As discussed above, this homogeneous 
component is absorbed into the phenomenological model parameters to avoid double-counting.  We 
therefore construct the generalized static susceptibility ${\bf \tilde X}(\bq)$ from the eigenstates 
of the bare Hamiltonian $\hat H_0$, defined in Eq.~(\ref{eq:bareham}).

Figure~\ref{fig:chi0} shows the largest eigenvalue $\chi(\bq)$ of  ${\bf \tilde X}(\bq)$
as a function of $\bq$ close to an instability approached upon cooling.
The generalized charge susceptibility allows transitions to charge-, bond-, and 
current-ordered phases, and the multi-peak structure in Fig.~\ref{fig:chi0} indicates proximity 
to more than one distinct ordered phase.
Because our supercell contains two primitive cells, the points
$\bq=(0,0)$ and $\bq=(\pi,\pi)$ are equivalent.  Furthermore, peaks at
$(q,q)$ and $(\pi-q,\pi-q)$ are related by symmetry.  There are,
therefore, only two distinct peaks in $\chi(\bq)$, corresponding to two
distinct phases. We use the notation $\bq_0 = (0,0)$ and $\bq_1 = (q_1,q_1)$ to denote 
these two kinds of peaks, while $\bq_2$ will be used later to denote peaks in
the axial direction at $(q_2,0)$ or $(0,q_2)$.

\begin{figure}
\includegraphics[width=0.45\textwidth]{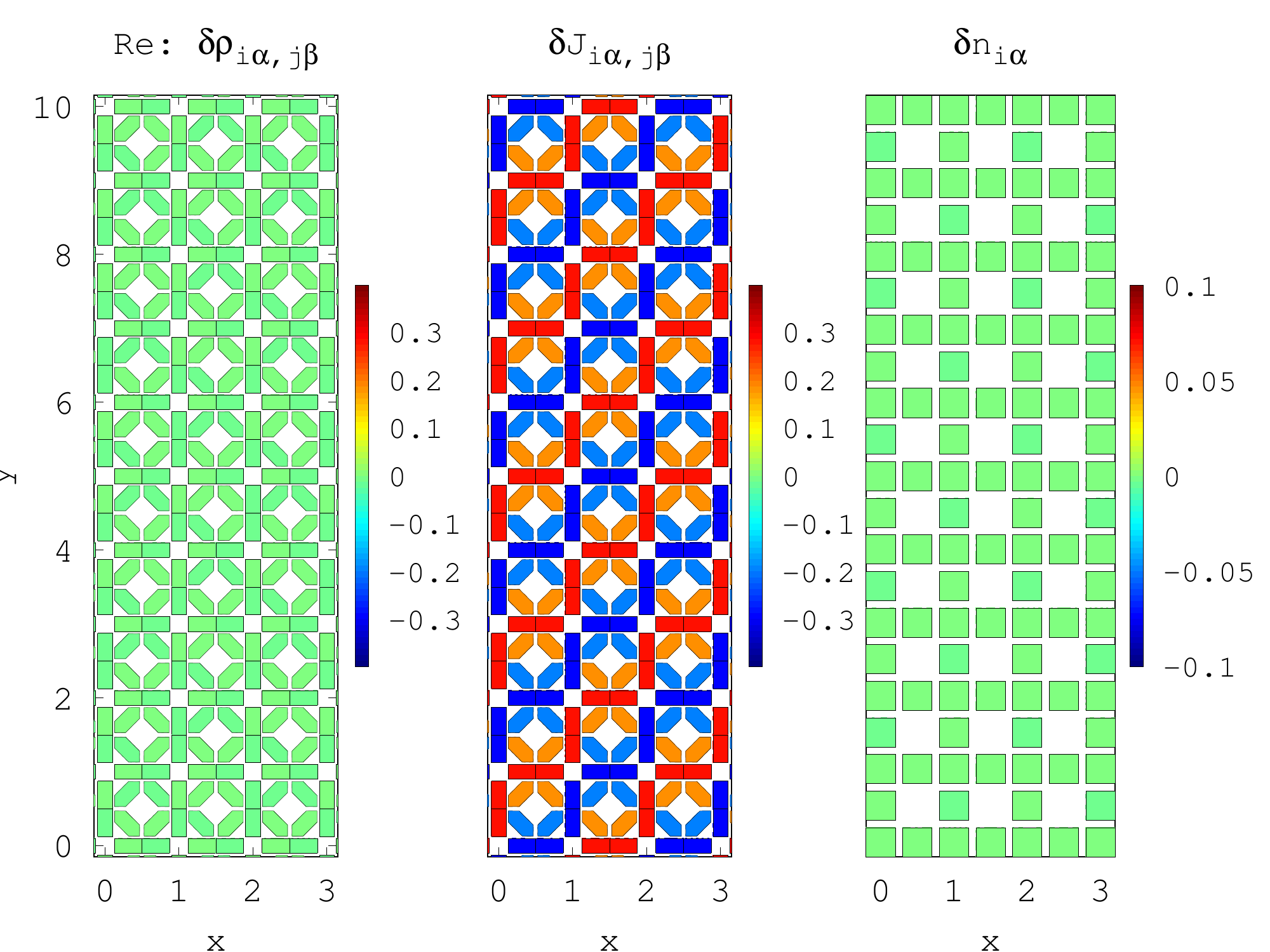}
\caption{(Color online) Components of the induced generalized charge density $\delta
  \rho_{i\alpha,j\beta}$ associated with the $\bq_0$ eigenmode for the
  susceptibility shown in Fig.~\ref{fig:chi0}.  (left) real part of
  $\delta \rho_{i\alpha,j\beta}$ for $\alpha \neq \beta$, (middle) induced currents
  $\delta J_{i\alpha,j\beta}$, (right) induced charge modulations $\delta
  n_{i\alpha} $.  Currents with a component of their
  flow in the positive $x$ direction are  deemed positive; 
  currents flowing entirely along the $y$ axis are
  positive in the positive $y$ direction.  $\delta
  \rho_{i\alpha,j\beta}$ is calculated from Eq.~(\ref{eq:drho}) with the prefactor
   $\chi(\bq)\varphi_\bq$ set  to one; the color scale is therefore arbitrary. }
\label{fig:rho0q0}
\end{figure}

For the chosen model parameters there are pronounced peaks at both $\bq_0$ and $\bq_1$.  
The peak at $\bq_0$ diverges first as $T$ is lowered, and is therefore the leading 
instability.  To determine the nature of the instability, we construct the  generalized 
charge density $\delta \rho_{i\alpha,j\beta}$  induced by an infinitesimally weak field
using Eq.~(\ref{eq:drho}). The left panel of Fig.~\ref{fig:rho0q0}(a) shows the real part of 
$\delta \rho_{i\alpha,j\beta}$ for $\alpha\neq \beta$,
%$i\neq j$, 
which is related by Eq.~(\ref{eq:PVrho}) to the bond-strength renormalization. The 
imaginary parts of $\delta \rho_{i\alpha,j\beta}$ are proportional to the  bond currents
$\delta J_{i\alpha,j\beta}$, which are shown in the middle panel of Fig.~\ref{fig:rho0q0}(b), while 
the orbital charge modulations $\delta n_{i\alpha} = \delta \rho_{i\alpha,i\alpha}$ are shown
in the right panel. From the figure, it is
apparent that the $\bq_0$ divergence corresponds to the onset of a
staggered loop-current pattern, with no associated charge or bond
order. (Note that $\bq_0$ is a {\em supercell} wavevector, and that
the current pattern has wavevector $(\pi,\pi)$ in terms of the primitive unit
cell.)  This is the same $\pi$LC pattern that was identified previously in 
Ref.~\onlinecite{Bulut:2015}.  

\begin{figure}
\includegraphics[width=\columnwidth]{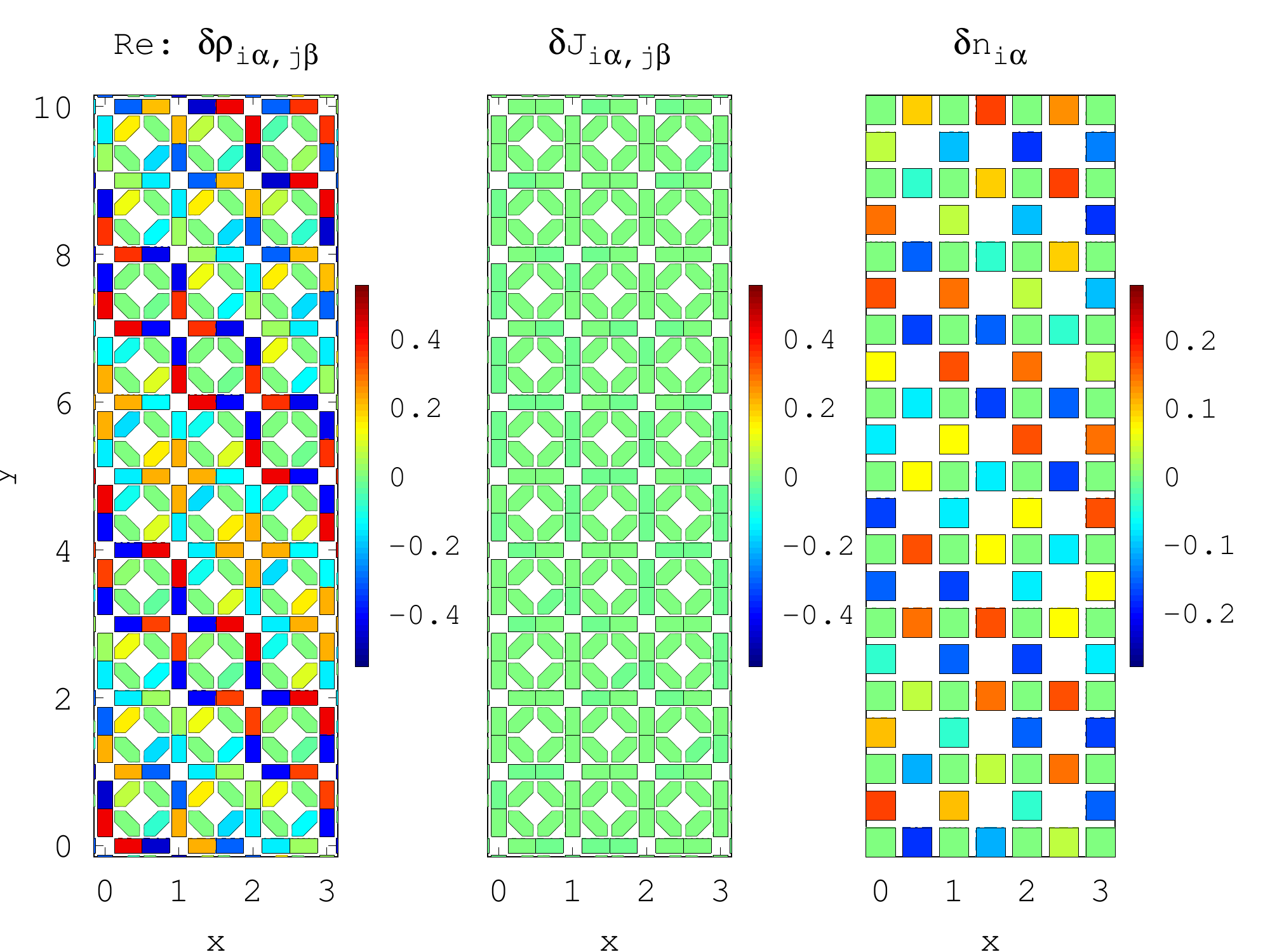}
\caption{(Color online) As in Fig.~\ref{fig:rho0q0}, but for the peak in
  Fig.~\ref{fig:chi0} at $\bq_1 = (0.84,0.84)$. This case corresponds to
  a diagonal dCDW with no circulating currents.  }
\label{fig:rho0q1}
\end{figure}

In contrast, Fig.~\ref{fig:rho0q1} shows that the subdominant peak at
$\bq_1=(0.84,0.84)$ corresponds to a diagonal dCDW with vanishing
orbital currents.  The period of this modulation is
$2\pi/(0.84\sqrt{2})=5.3$ primitive unit cells, similar to what is found
elsewhere, and agrees with the shortest wavevector which connects Fermi surface 
hotspots. This type of instability has been discussed at length in the 
literature\cite{Metlitski:PRB2010,Metlitski:2010vf,Holder:2012ks,Husemann:2012vg,YamasePRB2012,Efetov:2013,Bulut:2013,Meier:2013,Sau:2013vw}.

\begin{figure}
\includegraphics[width=\columnwidth]{{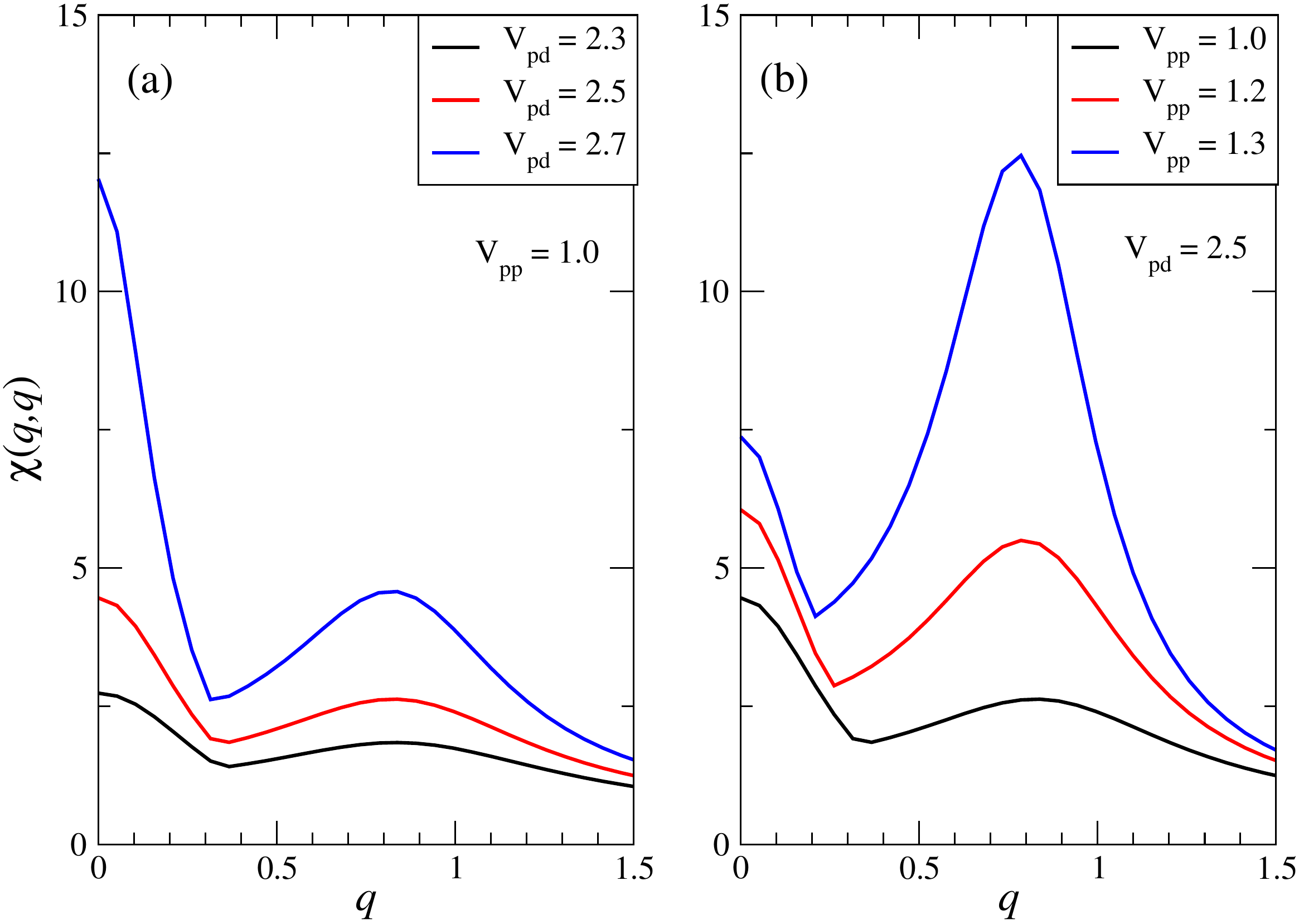}}
\caption{(Color online) Effects of $V_{pd}$ and $V_{pp}$ on the susceptibility eigenvalue $\chi(\bq)$.
Plots show cuts along the Brillouin zone diagonal, $\bq = (q,q)$. Results are for (a)  
fixed $V_{pp}=1.0$ and varying $V_{pd}$; (b) fixed $V_{pd}=2.5$ and varying $V_{pp}$. 
$V_{pd}$ enhances the loop-current susceptibility at $\bq_0$, while $V_{pp}$ enhances the 
charge susceptibility peak at $\bq_1$. Results are for $T=0.025$ and hole density $p=0.12$.}
\label{fig:chi_v_V}
\end{figure}

While the details of the competition between the $\pi$LC and charge ordered phases depend on the 
band structure, a simple picture emerges concerning the interactions driving these two phases.  
In Fig.~\ref{fig:chi_v_V} $\chi(\bq)$ is plotted along the Brillouin zone diagonal 
as functions of both $V_{pd}$ and $V_{pp}$:
Fig.~\ref{fig:chi_v_V}(a) shows that $\chi(\bq_0)$ is enhanced by increasing $V_{pd}$ while
Fig.~\ref{fig:chi_v_V}(b) shows that  $\chi(\bq_1)$  is enhanced by increasing $V_{pp}$.  
%Figs.~\ref{fig:chi_v_V}(a)-(c) show that $\chi(\bq_0)$ is enhanced by increasing $V_{pd}$ while
%Figs.~\ref{fig:chi_v_V}(d)-(f) show that  $\chi(\bq_1)$  is enhanced by increasing $V_{pp}$.  
This demonstrates that $V_{pd}$ drives the $\pi$LC phase while $V_{pp}$ drives the dCDW.

\begin{figure}
\includegraphics[width=\columnwidth]{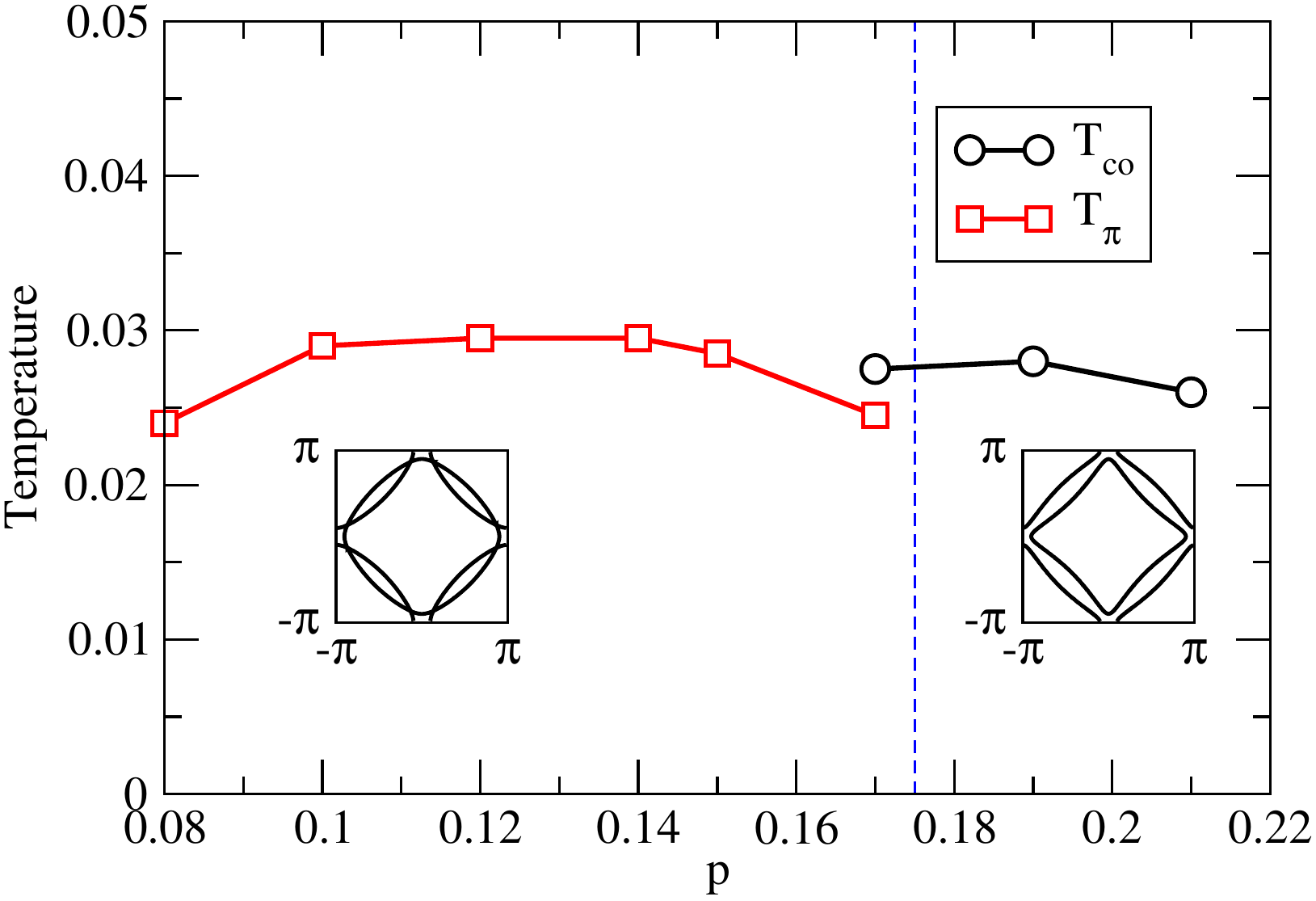}
\caption{(Color online) Phase diagram of the three-band model from linear response theory.  The
figure shows the doping dependence of the leading instability temperature of the generalized 
susceptibility ${\bf \tilde X}(\bq)$.  The van Hove hole density $p_\mathrm{vH}$ is indicated by
a vertical dashed line, and representative Fermi surfaces for $p < p_\mathrm{vH}$ and
$p > p_\mathrm{vH}$  are shown in the insets. The leading charge instability is towards
a diagonal dCDW for $p=0.17$, and to a  translationally invariant nematic phase 
consisting of an  intra-unit cell transfer of charge between
adjacent oxygen orbitals for $p>p_\mathrm{vH}$.\cite{Bulut:2013} 
Results are for $V_{pd} = 3.0$, $V_{pp} = 1.0$. }
\label{fig:phasediag}
\end{figure}

\begin{figure}
\includegraphics[width=\columnwidth]{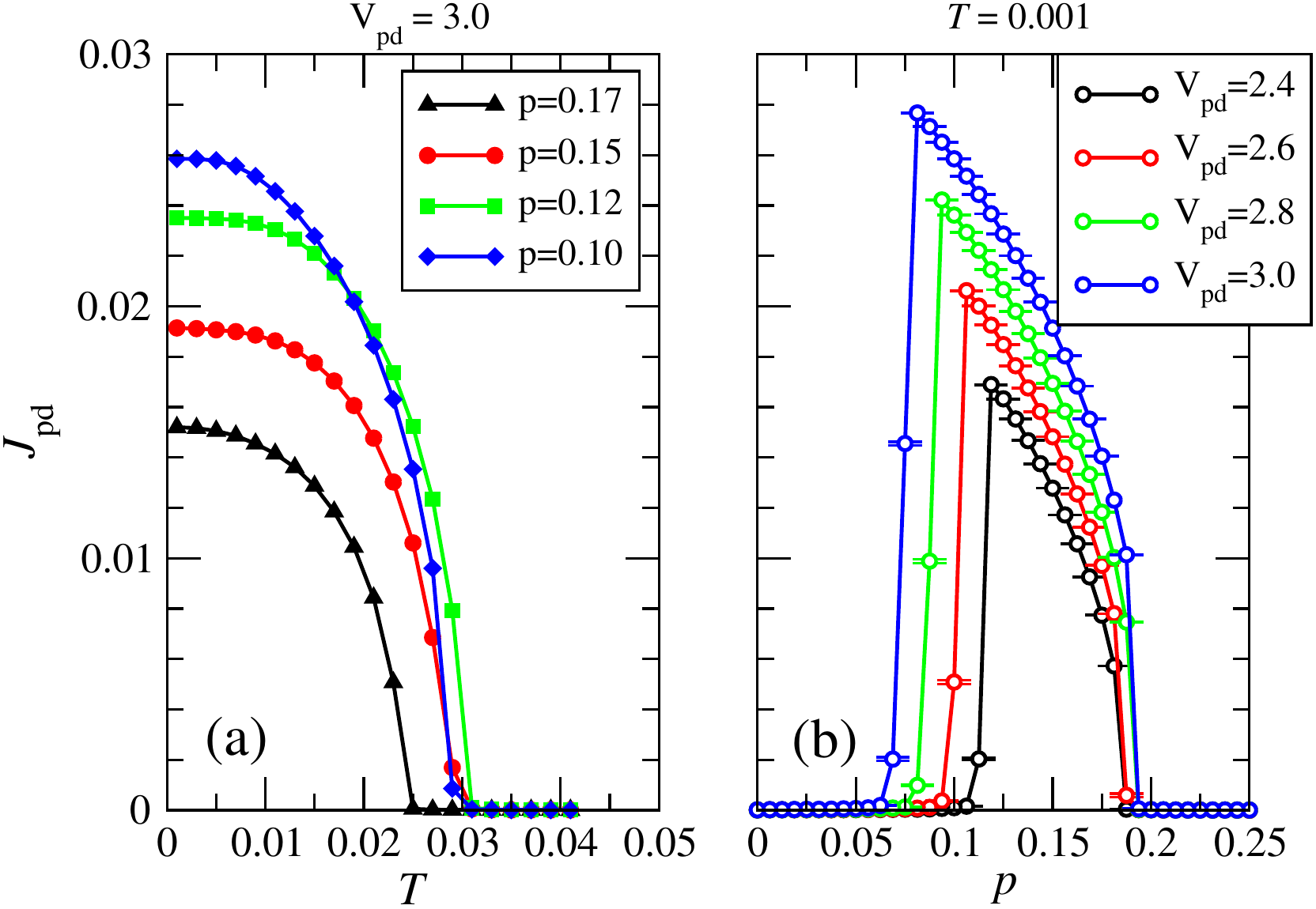}
\caption{(Color online) Self-consistent Hartree-Fock results for the orbital current as a 
  function of hole filling $p$, temperature $T$, and interaction $V_{pd}$.  The
  current $J_{pd}$ along the $p$-$d$ bond is shown as a function of
  (a) $T$ for $V_{pd} = 3.0$ and (b) $p$ for various $V_{pd}$ at
  $T=0.001$.  The ratio of the $p$-$p$ and $p$-$d$ bond currents is
  $J_{pp}/J_{pd} = 0.32$, independent of $T$, $p$, and $V_{pd}$.  It
  was previously found to depend on the ratio of
  $t_{pp}/t_{pd}$.\cite{Bulut:2015}
  The current is in units of $e t_{pd}/\hbar$, with $e$ the electron charge.
  }
\label{fig:Ppd}
\end{figure}

Figures~\ref{fig:phasediag} and \ref{fig:Ppd} show the dependence of the $\pi$LC phase on 
various model parameters. We caution that factors not included in our calculations must inevitably 
affect the phase diagram quantitatively. Notably, strong correlations renormalize the electronic 
effective mass, which grows as the hole doping $p$ is reduced, and the enhanced spin fluctuations 
make a further doping-dependent contribution to the self-energy.

Figure~\ref{fig:phasediag} shows the phase diagram which follows from the susceptibility 
calculations within the symmetry-unbroken normal state. This figure illustrates the particular 
significance of the van Hove filling $p_\mathrm{vH}$, which denotes the crossover from a hole-like
Fermi surface at $p<p_\mathrm{vH}$ to an electron-like Fermi surface at $p > p_\mathrm{vH}$.  
It was found previously\cite{Bulut:2013} that in the region $p<p_\mathrm{vH}$, the leading charge 
instability is to a diagonal dCDW, while for $p>p_\mathrm{vH}$ the tendency is towards either a 
$\bq=(0,0)$ nematic phase with an intra-unit cell charge redistribution or an axial dCDW.  
Figure~\ref{fig:phasediag} shows that the $\pi$LC phase is restricted to the region 
$p<p_\mathrm{vH}$, where it competes with the diagonal dCDW phase.

This result is confirmed by self-consistent HF calculations of the $\pi$LC phase diagram
in Fig.~\ref{fig:Ppd}. For these calculations, the self-energy
has the periodicity of the supercell, and Eq.~(\ref{eq:Ptilde}) can be
expressed simply in terms of the eigenvalues $E_{n\bk}$ and
eigenfunctions $\Psi_{\alpha n}(\bk)$ of the HF Hamiltonian,
$\hat H_{HF} = \hat H_0 + \hat V_{HF}$.  The $\bq={\bf 0}$ self-energy for bond $\ell$ is
\begin{eqnarray}
\tilde P^\ell &=& \sum_{\ell'} \tilde V^{\ell\ell'} 
 \frac 1N\sum_{\alpha\beta\bk'} 
g^{\ell'}_{\alpha\beta}(\bk^\prime)^\ast \Psi^\ast_{\beta n}(\bk')\Psi_{\alpha n}(\bk') f(E_{n\bk'}). 
\nonumber \\
\label{eq:Ptilde2}
\end{eqnarray}
Because the real part of Eq.~(\ref{eq:Ptilde2}) yields a homogeneous shift of the model
parameters, we have retained only the imaginary part of $\tilde P^\ell$ in the 
self-consistency cycle.  

Figure~\ref{fig:Ppd} shows the amplitude of the current $J_{pd}$ along the $p$-$d$ bonds in the
$\pi$LC phase. 
 The current is measured in units of $et_{pd}/\hbar$, so $J_{pd} = 0.01$ corresponds to a current of $\sim 1$ $\mu$A if $t_{pd} = 500$ meV.
The current sets in at $p_{vH}$ and its amplitude grows as hole doping is further 
reduced. The termination of the  $\pi$LC phase at $p \approx p_\mathrm{vH}$ is robust, as it is 
nearly independent of $V_{pd}$, and it is generally consistent with a recent experimental 
conclusion that the pseudogap phase is bounded by a Lifshitz transition.\cite{Benhabib:2015jd}   
However, the $p$-dependence of $J_{pd}$ is expected to be affected by strong correlations. In 
mean-field theory, the spectral gap associated with the $\pi$LC phase is proportional to the 
current amplitude. The HF self-energy Eq.~(\ref{eq:PVrho}) on the $p$-$d$ bonds, which determines 
both the spectral gap and $T_\pi$, is proportional to $V_{pd}$ and the generalized density 
$\rho_{pd}$ between $p$ and $d$ orbitals, while the current in Eq.~(\ref{eq:Jrho}) is proportional 
to $t_{pd}$ and $\rho_{pd}$. In the simplest picture, $t_{pd} \propto p$ so that the loop current 
amplitudes are renormalized downwards by strong correlations relative to the HF self-energy. This is 
similar to an effect predicted for strongly correlated superconductors:  in conventional 
superconductors, the superconducting $T_c$ is proportional to the superconducting gap $\Delta$; 
however, superfluid stiffness, and therefore $T_c$, is strongly reduced by strong correlations 
while the pairing gap remains large.\cite{Anderson:1987,Zhang:1988,Gros:1988,Paramekanti:2004,Haule:2007}  This suggests that the trends shown in 
Fig.~\ref{fig:Ppd}(b) qualitatively capture the spectral gap but not the LC amplitude.

The $\pi$LC phase stops abruptly at low $p$ at a value that does depend on $V_{pd}$; such a 
lower bound is not seen experimentally; however, the low-doping region of the phase diagram is 
complicated by strong correlations, the onset of a spin-glass phase, and by 
disorder\cite{Alvarez2008,Atkinson:2007dt} which are beyond the scope of our current calculations.

\subsection{Charge Instabilities in the Loop-Current State}
\label{sec:LC}
\begin{figure}
\includegraphics[width=0.45\textwidth]{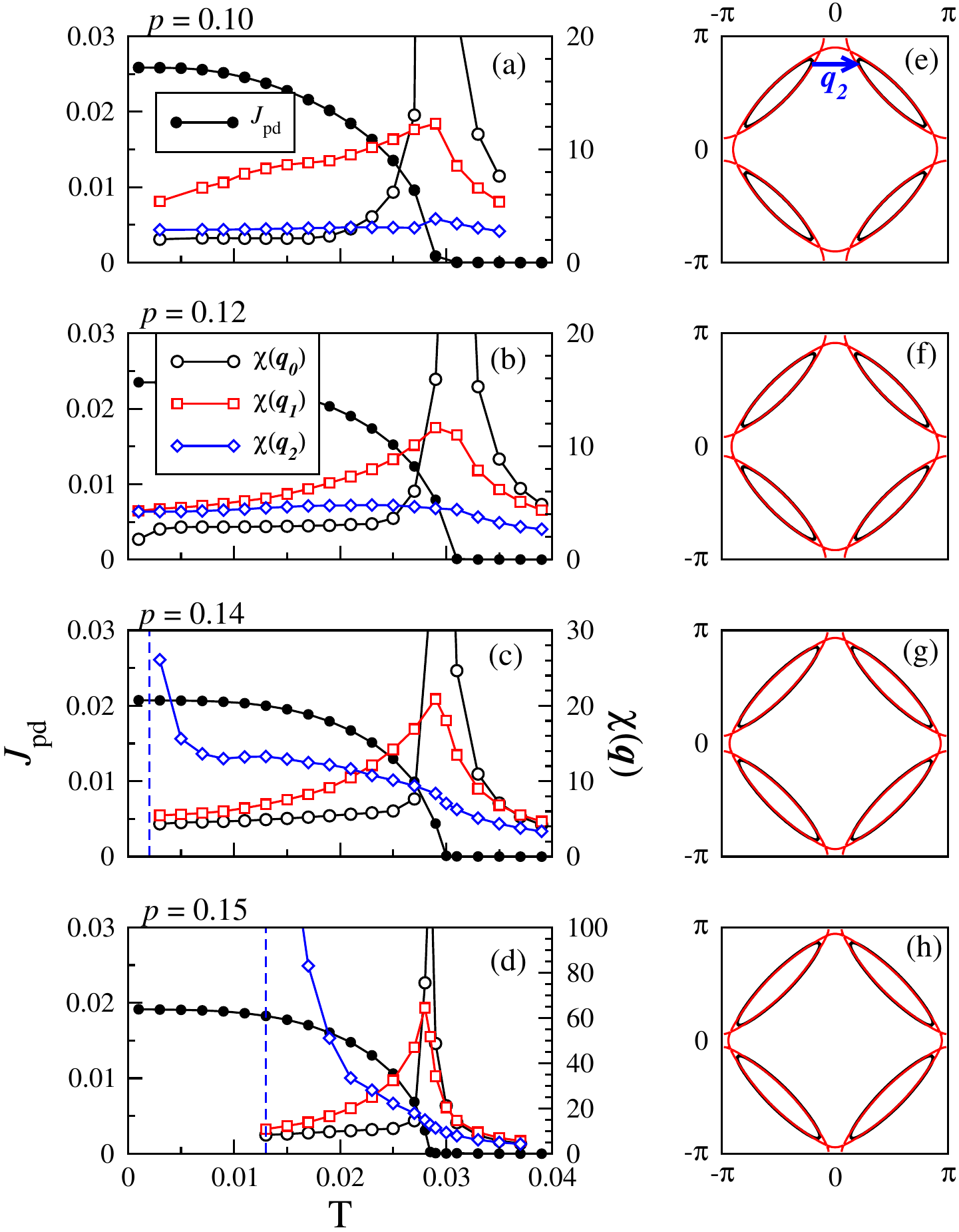}
\caption{(Color online) Temperature evolution of the peaks in $\chi({\bf q})$ at $\bq_0$, 
  $\bq_1$, and $\bq_2$ at four different fillings: (a) $p=0.10$, (b) $p=0.12$,
  (c) $p=0.14$, (d) $p=0.15$  The corresponding mean-field current along 
  the $p$-$d$ bond is also shown in each figure (left scale).
  The three wavevectors are $\bq_0 = (0,0)$, $\bq_1 = (q_1,q_1)$,
  $\bq_2 = (q_2,0)$, where $q_1$ and $q_2$ are the peak positions
  along diagonal and axial directions.  Fermi surfaces corresponding to the 
  different hole fillings are shown in (e)-(h) for the normal state (red lines) and for
  the $\pi$LC state (black lines). As shown in (e), the axial wavevector connects tips of the hole   pockets in the $\pi$LC phase.}
\label{fig:qs}
\end{figure}

To determine the leading instability within the $\pi$LC phase, we plot the
$T$-dependence of the leading eigenvalue $\chi(\bq)$ in Fig.~\ref{fig:qs} at $\bq_0$ (loop
current), $\bq_1$ (diagonal dCDW), and $\bq_2$ (axial dCDW). The susceptibility and its 
eigenvalues are now calculated using the self-consistent HF Hamiltonian for the $\pi$LC phase.
We focus on the region $p<p_\mathrm{vH}$, where loop currents are found, and 
results are shown at five different dopings between $p=0.10$ and $p=0.17$.
For reference, the Fermi surface and $T$-dependence of $J_{pd}$  are
also shown for each doping.

At temperatures above $T_\pi$, $\chi(\bq)$ grows at all three $\bq$ values as
$T$ is reduced.  For $p\leq 0.15$, $\chi(\bq_0)$ diverges first, signaling the onset of 
the $\pi$LC phase at $T_\pi$;  $\chi(\bq_0)$ then collapses rapidly in the ordered phase 
below $T_\pi$.  For all hole densities in Fig.~\ref{fig:qs}, the subleading peak is at $\bq_1$ 
for $T>T_\pi$, indicating a tendency towards a diagonal dCDW. This peak at ${\bf q}_1$ is reduced 
by the onset of loop currents, however, which demonstrates a strong competition between diagonal 
dCDW  and $\pi$LC order.

\begin{figure}
\includegraphics[width=\columnwidth]{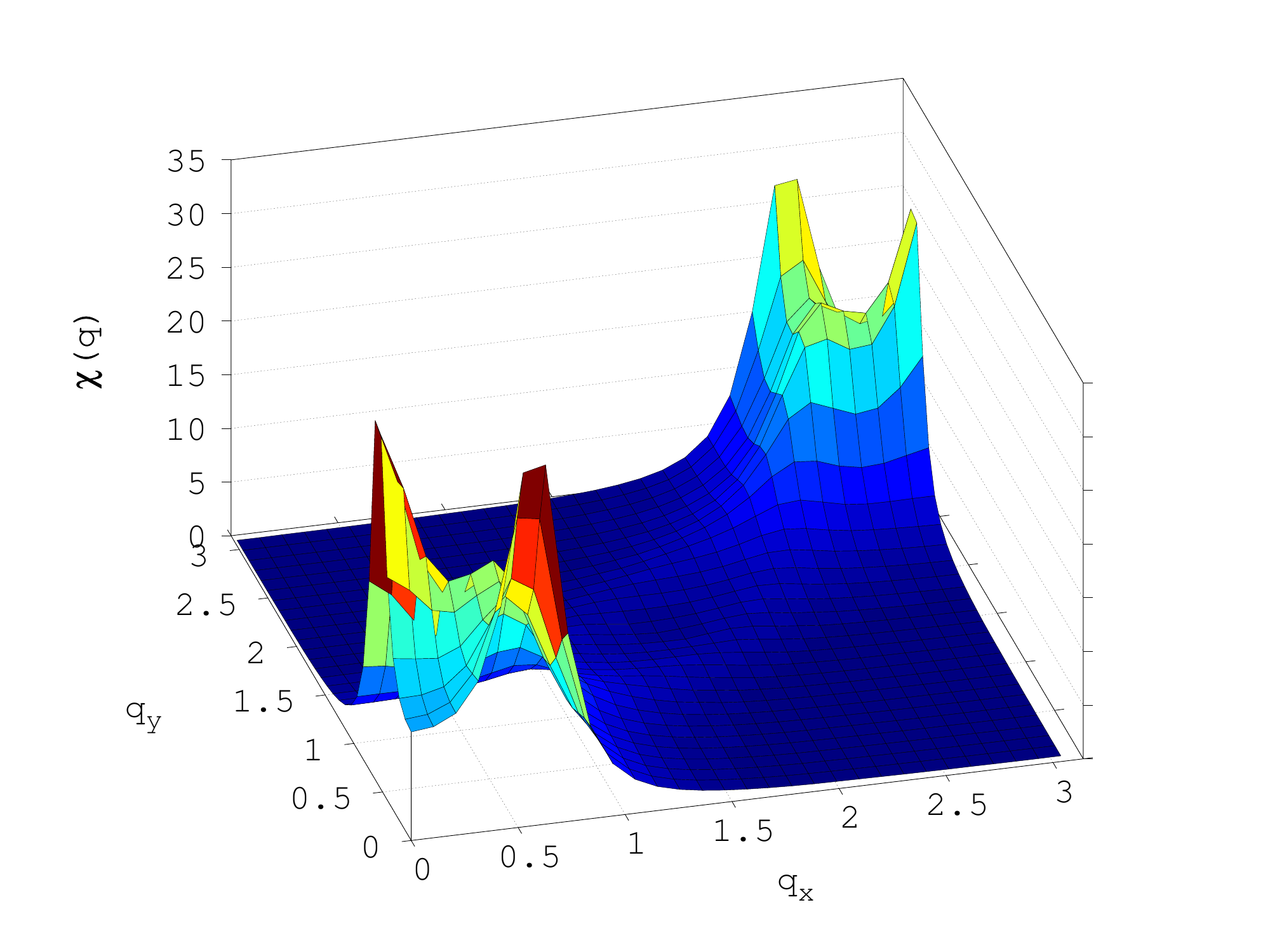}
\caption{ (Color online) Largest eigenvalue of the susceptibility matrix in the $\pi$LC
  phase.  Results are at $p=0.15$ for $V_{pd} = 3.0$, $V_{pp} = 1.0$,
  and at $T=0.021$. }
\label{fig:chi_p15}
\end{figure}

\begin{figure}
\includegraphics[width=\columnwidth]{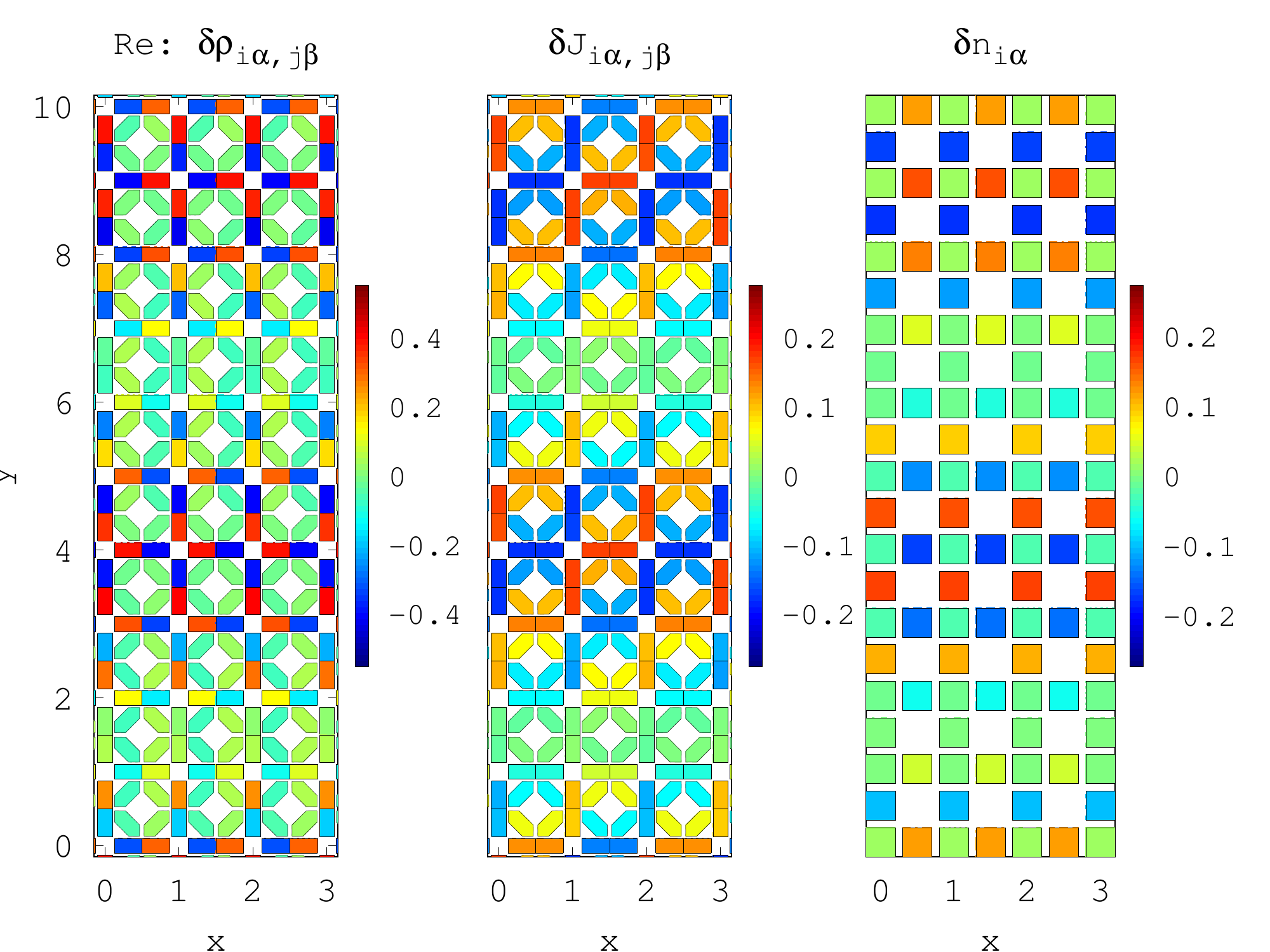}
\caption{ (Color online) Left panel: induced shift in the real part of $\langle c^\dagger_{i\alpha}
  c_{j\beta}\rangle$, middle panel: the bond current, and left panel: the charge
  density.  Parameters are as in Fig.~\ref{fig:chi_p15}. The charge
  modulations on the Cu sites are $\sim 15\%$ of those on the O sites.}
\label{fig:order_p15}
\end{figure}

In contrast, there is only a weak competition between axial dCDW and $\pi$LC order.
Above $T_\pi$, $\chi(\bq_2)$ has positive curvature characteristic of growth towards a divergence;
however, all curves show an inflection point slightly below $T_\pi$ indicating that the onset of 
loop currents interrupts this divergence.  
%The competition between the phases is relatively weak, and
%is the smallest of the three susceptibility eigenvalues;  however,  below $T_\pi$ 
%the axial susceptibility is much less affected by $\pi$LC order than is the diagonal 
%susceptibility.
Rather than being suppressed by loop currents, $\chi(\bq_2)$ tends to saturate below $T_\pi$ at a constant value  [Figs.~\ref{fig:qs}(a) and (b)],  which can be an order of magnitude larger than at high $T$.  At some doping levels [Figs.~\ref{fig:qs}(c) 
and (d)], $\chi(\bq_2)$ actually diverges below $T_\pi$, signaling the onset of an axially 
oriented dCDW. This is shown for  $p=0.15$ in Fig.~\ref{fig:chi_p15}, which shows the emergence 
of strong peaks at $\bq_2 = (q_2,0)$ and symmetry related points. The corresponding eigenmode is 
illustrated in Fig.~\ref{fig:order_p15}: there is a pronounced transfer of charge between O$p$ 
orbitals, with an amplitude that is modulated along the $y$-axis 
[right panel of Fig.~\ref{fig:order_p15}]. There is a smaller charge modulation on 
the Cu sites, amounting to $\sim 15\%$ of the O$p$ modulations. This is similar to the axial dCDW found previously for a 
phenomenological pseudogap model\cite{Atkinson:2014}, and both the ordering wavevector and 
$d$-wave like form factor of the charge modulations are consistent with 
experiments.\cite{Comin:2014vq,Fujita:2014kg}

Concomitantly, the real part of $\delta \rho_{i\alpha,j\beta}$ (left panel in Fig.~\ref{fig:order_p15})
inhomogeneously modulates the effective hopping strength, while the imaginary part corresponds 
to an incommensurate modulation of the bond current (middle panel in Fig.~\ref{fig:order_p15}).  
We have checked that this incommensurate current pattern conserves charge at each vertex of the 
lattice. Indeed, it is straightforward to construct such a modulated current pattern by hand by 
requiring that current is conserved at each vertex. Current conservation at the  Cu site at 
position $(m,n)a_0$, where $a_0$ is the primitive lattice constant, requires that the current 
coming in along the $x$-axis must be carried out along the $y$-axis, namely 
\begin{equation}
\sum_{s=\pm 1}\left[ I_x(m+\frac s2,n) + I_y(m,n+\frac s2)\right] =0\, .
\label{eq:Isum}
\end{equation}
For a modulation wavevector $\bq_2 = (0,q_2)$, this constraint implies,
\begin{equation}
I_x(m+\frac 12, n) = I_0 (-1)^{n+m} \cos(q_2na_0+\theta),
\label{eq:Ix}
\end{equation}
where $\theta$ is an arbitrary constant phase, and 
\begin{equation}
I_y(m,n+\frac 12) = -I_0 (-1)^{n+m} \frac{\cos[q_2(n+\frac 12)a_0+\theta]}{\cos (q_2a_0/2)}.
\label{eq:Iy}
\end{equation}
Current conservation along the oxygen-oxygen bonds is simpler, as it requires only that the 
current is constant around each loop within a plaquette.

The weak competition between axial dCDW and $\pi$LC order has implications for the role 
of disorder.  While the growth of critical diagonal dCDW fluctuations is interrupted at
$T_\pi$, $\chi(\bq_2)$ in some instances saturates below $T_\pi$ at values
that are substantially enhanced relative to the noninteracting case
(where $\chi(\bq)$ is a number of order 1).  Figure~\ref{fig:qs}(c), for example, is
characterized by a wide temperature range below $T_\pi$ where $\chi(\bq_2)$ is more than an order of magnitude larger than in the noninteracting case. 
In this case crystalline disorder, for example due to dopant atoms, will induce
short-range charge correlations with a strong $\bq_2$ component even well above the charge-ordering transition at $T_\mathrm{co}=0.002$.

This is consistent with what is observed in the cuprates, where
static  short-range dCDW correlations develop at temperatures as high as $\sim 150$ K;\cite{Ghiringhelli:2012bw,Chang:2012vf} and true long-range dCDW order (with correlation lengths large enough to observe magneto-oscillation effects) only occurs at much lower temperatures, of order $\sim 50$ K.\cite{Wu:2011ke}

Finally, we note that the generalized susceptibility diverges simultaneously at symmetry-related points along the $x$- and $y$-directions  (Fig.~\ref{fig:chi_p15}).  Our linearized equations cannot determine whether uniaxial order, with ordering wavector $(q_2,0)$ or $(0,q_2)$, 
or biaxial (checkerboard) order in which both Fourier peaks are simultaneously present, is energetically preferred.  Experimentally, domains of uniaxial order are seen in Bi$_2$Sr$_2$CaCu$_2$O$_{8+x}$,\cite{Fujita:2014kg}  while biaxial order is implied by magneto-oscillation experiments in YBa$_2$Cu$_3$O$_{6+x}$.\cite{Sebastian:2012wh} 

In our calculations, the biaxial dCDW state is of particular interest because it breaks all mirror symmetries of the lattice, and coupled with the time-reversal symmetry breaking of the $\pi$LC phase should generate a polar Kerr effect,\cite{Wang:2014it,Gradhand:2015fr}  similar to what has been measured in both YBa$_2$Cu$_3$O$_{6+x}$\cite{Xia:2008cc} and Pb$_{0.55}$Bi$_{1.5}$Sr$_{1.6}$La$_{0.4}$CuO$_{6+\delta}$.\cite{He:2011}  This mechanism should be distinguished from other proposals involving microscopic currents:  in Refs.~\onlinecite{Wang:2014it} and \onlinecite{Gradhand:2015fr}, the currents run along the edges of charge stripes, while in Ref.~\onlinecite{Sharma:2015} a combination of staggered loop currents and $d_{xy}$ bond order is proposed to explain the polar Kerr measurements.  It remains unclear whether nanodomains of uniaxial order might also lead to a polar Kerr effect in our model.
 
\section{Conclusions}
Motivated by the recent discovery of ubiquitous charge order within the pseudogap phase of underdoped cuprate superconductors, we have studied the development of $d$-charge density waves from within a pseudogap phase generated by a staggered $\pi$-loop current.     Our main finding is that the $\pi$LC phase competes strongly with the dominant diagonal dCDW phase, and may weaken it sufficiently that axial dCDW order emerges as the leading charge instability.  The resulting charge structure is consistent with x-ray scattering and STM experiments.  A unique feature of the coexistence of dCDW and $\pi$LC order is the emergence of an incommensurate modulation of the loop current amplitude, illustrated in Fig.~\ref{fig:order_p15}.  If the dCDW has a checkerboard structure, then the resulting incommensurate current pattern breaks both mirror and time-reversal symmetries and should generate a polar Kerr effect.

\section*{Acknowledgments}
A.P.K.\ and S.B.\ were supported by the Deutsche Forschungsgemeinschaft through TRR80. 
W.A.A.\ acknowledges support by the National Sciences and Engineering Research 
Council (NSERC) of Canada.

\bibliography{RPA}

\appendix

\section{Hartree-Fock Approximation and Basis Functions}
\label{app:A}
In this appendix, we discuss technical details of our treatment
of the interactions in the Hartree-Fock approximation.  
First, we give an explicit form for $V_{\alpha\beta}(\bq)$, defined by
Eq.~(\ref{eq:Vabq}). Referring to Fig.~\ref{fig:unitcell}, we obtain
\begin{eqnarray} 
V_{12}(\bq) &=& V_{45}(\bq) = V_{24}(\bq) = V_{51}(\bq) = V_{pd} e^{iq_y/2} \nonumber \\
V_{21}(\bq) &=& V_{54}(\bq) = V_{42}(\bq) = V_{15}(\bq) = V_{pd} e^{-iq_y/2} \nonumber \\
V_{13}(\bq) &=& V_{34}(\bq) = V_{46}(\bq) = V_{61}(\bq) = V_{pd} e^{iq_x/2} \nonumber \\
V_{31}(\bq) &=& V_{43}(\bq) = V_{64}(\bq) = V_{16}(\bq) = V_{pd} e^{-iq_x/2} \nonumber \\
V_{23}(\bq) &=& V_{32}(\bq) = V_{56}(\bq) = V_{65}(\bq) \nonumber \\
&=& 2 V_{pp} \cos\left( \frac{q_x-q_y}2 \right ) \nonumber \\
V_{26}(\bq) &=& V_{62}(\bq) = V_{53}(\bq) = V_{35}(\bq) \nonumber \\ 
&=& 2 V_{pp} \cos\left( \frac{q_x+q_y}2 \right ), 
\end{eqnarray}
with all other matrix elements zero.

The correspondence between the basis function $\ell$ and the orbital label is given in
Table~\ref{tab:P}.  From this table, we learn for example that
\begin{equation}
g^5_{\alpha\beta}(\bk) = e^{ik_y/2} \delta_{\alpha,4} \delta_{\beta,5}.
\end{equation}
For this basis,
\begin{equation}
\tilde V^{\ell\ell'}(\bq) = \left \{ \begin{array}{ll}
-V_{pd} & \ell = \ell'; \ell \in [1,16] \\
-V_{pp} & \ell = \ell'; \ell \in [17,32] \\
U_d & \ell = \ell' = 33, 36 \\
U_p & \ell = \ell' = 34, 35, 37, 38 \\
2V_{\alpha\beta}(\bq) & \ell \neq \ell'; \, \ell,\ell' \in [33,38] \\
0 & \mbox{otherwise}
\end{array} \right.
\end{equation}
For the matrix elements containing $V_{\alpha\beta}(\bq)$, $\ell$
determines $\alpha$ and $\ell'$ determines $\beta$, with the
connection given by Table~\ref{tab:P}.

\begin{table*}
\begin{tabular}{l|l|l|l||l|l|l|l||l|l|l|l|l||l|l|l|l|l||l|l|l|l}
\hline
 $\ell$ & $\alpha_\ell$ & $\beta_\ell$ & $g^\ell(\bk)$ & 
 $\ell$ & $\alpha_\ell$ & $\beta_\ell$ & $g^\ell(\bk)$ &
 $\ell$ & $\alpha_\ell$ & $\beta_\ell$ & $g^\ell(\bk)$ & ${\bf r}_{\alpha\rightarrow \beta}$ & 
 $\ell$ & $\alpha_\ell$ & $\beta_\ell$ & $g^\ell(\bk)$ & ${\bf r}_{\alpha\rightarrow \beta}$ &
 $\ell$ & $\alpha_\ell$ & $\beta_\ell$ & $g^\ell(\bk)$ 
\\
\hline 
\hline 
1 & 1 & 2 & $e^{ik_y/2}$ & 9 & 2 & 1 & $e^{-ik_y/2}$ & 17 & 2 & 3 &$e^{i(k_x-k_y)/2}$& $+{\bf a}_2/2$ & 25 & 3 & 2 &$e^{-i(k_x-k_y)/2}$& $-{\bf a}_2/2$ & 33 & 1 & 1 & 1 \\
2 & 1 & 3 & $e^{ik_x/2}$ & 10& 3 & 1 & $e^{-ik_x/2}$ & 18 & 2 & 3 &$e^{-i(k_x-k_y)/2}$& $-{\bf a}_2/2$ & 26 & 3 & 2 &$e^{i(k_x-k_y)/2}$& $+{\bf a}_2/2$ & 34 & 2 & 2 & 1 \\
3 & 1 & 5 & $e^{-ik_y/2}$ & 11& 5 & 1 &$e^{ik_y/2}$ & 19 & 2 & 6 &$e^{i(k_x+k_y)/2}$& $+{\bf a}_1/2$ & 27 & 6 & 2 &$e^{-i(k_x+k_y)/2}$& $-{\bf a}_1/2$ & 35 & 3 & 3 & 1 \\
4 & 1 & 6 & $e^{-ik_x/2}$ & 12& 6 & 1 &$e^{ik_x/2}$ &20 & 2 & 6 &$e^{-i(k_x+k_y)/2}$& $-{\bf a}_1/2$ & 28 & 6 & 2 &$e^{i(k_x+k_y)/2}$&$+{\bf a}_1/2$ & 36 & 4 & 4 & 1 \\
5 & 4 & 5 & $e^{ik_y/2}$  &13& 5 & 4 &$e^{-ik_y/2}$ &21 & 5 & 6 &$e^{i(k_x-k_y)/2}$& $+{\bf a}_2/2$  & 29 & 6 & 5 &$e^{-i(k_x-k_y)/2}$&$-{\bf a}_2/2$ & 37 & 5 & 5 & 1 \\
6 & 4 & 6 & $e^{ik_x/2}$ & 14& 6 & 4 &$e^{-ik_x/2}$ & 22 & 5 & 6 &$e^{-i(k_x-k_y)/2}$& $-{\bf a}_2/2$ & 30 & 6 & 5 &$e^{i(k_x-k_y)/2}$&$+{\bf a}_2/2$ & 38 & 6 & 6 & 1 \\
7 & 4 & 2 & $e^{-ik_y/2}$ & 15& 2 & 4 & $e^{ik_y/2}$ & 23 & 3 & 5 &$e^{i(k_x+k_y)/2}$& $+{\bf a}_1/2$ & 31 & 5 & 3 &$e^{-i(k_x+k_y)/2}$& $-{\bf a}_1/2$ &&&&\\
8 & 4 & 3 & $e^{-ik_x/2}$ & 16& 3 & 4 &$e^{ik_x/2}$ & 24 & 3 & 5 &$e^{-i(k_x+k_y)/2}$& $-{\bf a}_1/2$ & 32 & 5 & 3 &$e^{i(k_x+k_y)/2}$& $+{\bf a}_1/2$ &&&&\\
\hline
\end{tabular}
\caption{The basis functions 
$g^\ell_{\alpha\beta}(\bk) = g^\ell(\bk)\delta_{\alpha,\alpha_\ell}
\delta_{\beta,\beta_\ell}$.
  The index $\ell$ labels the different basis functions, and each
  $\ell$ corresponds to a unique pair of orbitals $\alpha_\ell$ and $\beta_\ell$
  for which $g^\ell_{\alpha_\ell \beta_\ell}(\bk)$ is nonzero.  In this basis,
  and for $\ell \in [1,32]$, the quantity $\tilde P^\ell$ defined in
  Eq.~(\ref{eq:Ptilde}) is simply the bond self-energy
  $P_{i\alpha,j\beta}$ for nearest neighbor sites $i\alpha$,
  $j\beta$.  For $\ell \in [17,32]$, there are two nearest-neighbor
  pairs for each $\alpha,\beta$, and to remove the ambiguity, the
  table shows the vector ${\bf r}_{\alpha\rightarrow \beta}$ pointing
  from $\alpha$ to $\beta$. The basis functions with $\ell \in
  [33,38]$ are used to represent the Hartree self-energies. }
\label{tab:P}
\end{table*}

\section{Symmetries of the Susceptibility}
\label{app:C}
\subsection{Hermiticity of $\mathbf{\tilde X}(\bq)$}
From Eq.~(\ref{eq:X0}), we obtain the relationship for the static susceptibility
\begin{equation}
\tilde X_0^{\ell\ell'}(\bq)^\ast = \tilde X_0^{\ell' \ell}(\bq),
\end{equation}
or, in matrix notation,
$\mathbf{\tilde X_0}(\bq)^\dagger = \mathbf{\tilde X_0}(\bq)$.
Since $\mathbf{\tilde V}(\bq)^\dagger = \mathbf{\tilde V}(\bq)$, it
also follows that
\begin{equation}
\mathbf{\tilde X}(\bq)^\dagger = \mathbf{\tilde X}(\bq).
\end{equation}

\subsection{Relation between $\mathbf{\tilde X}(\bq)$ and
$\mathbf{\tilde X}(-\bq)$}
 It also follows from Eq.~(\ref{eq:X0}) that 
\begin{equation}
\tilde X_0^{\ell\ell'}(-\bq) = \tilde X_0^{\ellbar'\ellbar}(\bq)
= \left [ \tilde X_0^{\ellbar \ellbar'}(\bq) \right ]^\ast.
\end{equation}
where $\ellbar$ represents the same bond as $\ell$, but oriented in
the opposite sense. Let ${\bf T}$ be the unitary
matrix that swaps bonds to the opposite orientation, we obtain the matrix representation
${\bf\tilde  X_0}(-\bq)  = {\bf T \tilde X_0}(\bq) {\bf T}^\dagger$.
Because
\begin{equation}
{\bf \tilde V}(\bq)  = {\bf T \tilde V}(\bq) {\bf T}^\dagger; \qquad
{\bf \tilde V}(-\bq)  = {\bf\tilde V}(\bq)^\ast
\end{equation}
it also happens that
\begin{equation}
{\bf\tilde  X}(-\bq)  = {\bf T \tilde X}(\bq)^\ast {\bf T}^\dagger.
\label{eq:Xq}
\end{equation}
For the labeling of the bonds shown in Table~\ref{tab:P},
\begin{equation}
{\bf T} = \left [ \begin{array}{cc|cc|c}
{\bf 0}_{8\times 8} & {\bf 1}_{8\times 8} 
& {\bf 0}_{8\times 8} & {\bf 0}_{8\times 8}  & {\bf 0}_{8\times 6}  \\
{\bf 1}_{8\times 8} & {\bf 0}_{8\times 8} 
& {\bf 0}_{8\times 8} & {\bf 0}_{8\times 8}  & {\bf 0}_{8\times 6}  \\
\hline
{\bf 0}_{8\times 8} & {\bf 0}_{8\times 8} 
& {\bf 0}_{8\times 8} & {\bf 1}_{8\times 8}  & {\bf 0}_{8\times 6}  \\
{\bf 0}_{8\times 8} & {\bf 0}_{8\times 8} 
& {\bf 1}_{8\times 8} & {\bf 0}_{8\times 8}  & {\bf 0}_{8\times 6}  \\
\hline
{\bf 0}_{6\times 8} & {\bf 0}_{6\times 8} 
& {\bf 0}_{6\times 8} & {\bf 0}_{6\times 8}  & {\bf 1}_{6\times 6}  
\end{array}\right ]
\label{eq:T}
\end{equation}
Note that ${\bf T}^\dagger = {\bf T}$ in this case.

\subsection{Eigenvectors of $\mathbf{\tilde X}(\bq)$}
Equation~(\ref{eq:Xq}) implies that the eigenvalue equation,
$\mathbf{ \tilde X}(\bq) {\bf v}_\bq = \chi(\bq) {\bf v}_\bq$,
transforms as
\begin{eqnarray}
\mathbf{ T \tilde X}(\bq)^\ast \mathbf{T^\dagger T} {\bf v}_\bq^\ast
& =& \chi(\bq) {\bf T v}_\bq^\ast \\
\mathbf{ \tilde X}(-\bq)\mathbf{T} {\bf v}_\bq^\ast
& =& \chi(\bq) {\bf T v}_\bq^\ast.
\end{eqnarray}
Because $\chi(-\bq) = \chi(\bq)$, it follows that ${\bf T
  v}_\bq^\ast = e^{i\theta} {\bf v}_{-\bq}$, where $\theta$ is an
arbitrary phase.  Without loss of generality, we take $\theta=0$ and 
\begin{equation}
{\bf T v}_\bq^\ast = {\bf v}_{-\bq}.
\label{eq:vtransform}
\end{equation}

\subsection{Simplification of the equation for $\delta\rho_{i\alpha,j\beta}$}

Equation (\ref{eq:rhoq1}) gives the induced generalized charge density
\begin{equation}
\delta \bm{\tilde \rho}(\bq) = 
-\mathbf{\tilde X}(-\bq)^\ast \bm{\tilde \phi}(-\bq)^\ast.
\label{eq:rhophi}
\end{equation}
Using the symmetry relations above along with the hermiticity condition (\ref{eq:hermiticity}),
Eq.~(\ref{eq:rhophi}) becomes
\begin{equation}
\bm{\tilde \rho}(\bq) 
%-{\bf \tilde X}(-\bq)^\ast \bm{\tilde \phi}(-\bq)^\ast 
= - {\bf T \tilde X}(\bq) \bm{\tilde \phi}(\bq).
\end{equation}
 Letting $\ell$ correspond to
the bond $(i\alpha,j\beta)$, 
\begin{eqnarray}
\delta \rho_{i\alpha,j\beta} &=& \delta  \tilde \rho^{\ell}(\bq) 
e^{i\bqt\cdot (\br_{i\alpha}+\br_{j\beta})} 
+ \delta \tilde \rho^{\ell}(-\bq) e^{-i\bqt\cdot (\br_{i\alpha}+\br_{j\beta})} \nonumber \\
&=& -e^{i\bqt\cdot (\br_{i\alpha}+\br_{j\beta})} 
\left [ {\bf T\tilde X(\bq) \tilde \phi(\bq) }\right ]^\ell 
\nonumber \\
&&- e^{-i\bqt\cdot (\br_{i\alpha}+\br_{j\beta})} \left [ 
 {\bf \tilde X(\bq)^\ast \tilde \phi(\bq)^\ast }\right ]^\ell.
\end{eqnarray}
If $\bm{\tilde \phi}(\bq)$ is proportional to an
eigenvector of ${\bf \tilde X}(\bq)$ with real eigenvalue $\chi(\bq)$,
then
\begin{eqnarray}
\delta \rho_{i\alpha,j\beta} &=& -\chi(\bq)\Big \{ 
e^{i\bqt\cdot (\br_{i\alpha}+\br_{j\beta})}  
\left [ {\bf T}\bm{ \tilde \phi}(\bq)\right ]^\ell 
\nonumber \\
&& 
+ e^{-i\bqt\cdot (\br_{i\alpha}+\br_{j\beta})} \left [ 
 \bm{ \tilde \phi}(\bq)^\ast \right ]^\ell 
\Big \} \nonumber \\
 &=& -\chi(\bq) \left \{ 
e^{i\bqt\cdot (\br_{i\alpha}+\br_{j\beta})}  
 \tilde \phi^\ellbar (\bq) \right . 
 \nonumber \\
&& 
\left .
+ e^{-i\bqt\cdot (\br_{i\alpha}+\br_{j\beta})} 
  \tilde \phi^\ell(\bq)^{\ast } \right \}. \nonumber\\
\label{eq:rhoij}
\end{eqnarray}

Taking $\tilde \phi^\ellbar(\bq) = \varphi_\bq v_\bq^\ellbar$, we obtain Eq.~(\ref{eq:drho2}).

\end{document}